\documentclass[aps,floats]{revtex4}
\usepackage{amsmath,amssymb}
\usepackage{graphicx,epsfig}

\begin{document}
\bibliographystyle {plain}

\def\oppropto{\mathop{\propto}} 
\def\opsimeq{\mathop{\simeq}}
\def\opoverderline{\mathop{\overline}}
\def\operarrow{\mathop{\longrightarrow}}
\def\opsim{\mathop{\sim}} 
\def\oplim{\mathop{\lim}} 

\def\fig#1#2{\includegraphics[height=#1]{#2}}
\def\figx#1#2{\includegraphics[width=#1]{#2}}


\title{A critical Dyson hierarchical model for the Anderson localization transition } 


 \author{ C\'ecile Monthus and Thomas Garel }
  \affiliation{ Institut de Physique Th\'{e}orique, CNRS and CEA Saclay,
 91191 Gif-sur-Yvette, France}

\begin{abstract}
A Dyson hierarchical model for Anderson localization, containing non-random hierarchical hoppings and random on-site energies, has been studied in the mathematical literature since its introduction by Bovier [J. Stat. Phys. 59, 745 (1990)], with the conclusion that this model is always in the localized phase. Here we show that if one introduces alternating signs in the hoppings along the hierarchy (instead of choosing all hoppings of the same sign), it is possible to reach an Anderson localization critical point presenting multifractal eigenfunctions and intermediate spectral statistics. The advantage of this model is that one can write exact renormalization equations for some observables. In particular, we obtain that the renormalized on-site energies have the Cauchy distributions for exact fixed points. Another output of this renormalization analysis is that the typical exponent of critical eigenfunctions is always $\alpha_{typ}=2$, independently of the disorder strength. We present numerical results concerning the whole multifractal spectrum $f(\alpha)$ and the compressibility $\chi$ of the level statistics, both for the box and the Cauchy distributions of the random on-site energies. We discuss the similarities and differences with the ensemble of ultrametric random matrices introduced recently by Fyodorov, Ossipov and Rodriguez [J. Stat. Mech. L12001 (2009)].

\end{abstract}

\maketitle

 \section{ Introduction} 

To better understand the notion of phase transition in statistical physics,
Dyson \cite{dyson} has introduced long ago a hierarchical ferromagnetic spin model,
which can be studied via exact renormalization for probability distributions.
Since the hierarchical couplings correspond to long-ranged power-law couplings in real space,
phase transitions are possible already in one dimension.
This type of hierarchical model has thus attracted a great interest 
in statistical physics, both among mathematicians
\cite{bleher,gallavotti,book,jona} and among physicists \cite{baker,mcguire,Kim}.
In the field of quenched disordered models, hierarchical models have also been 
introduced for spin systems with random fields \cite{randomfield}
or with random couplings \cite{sgdysonAT,sgdysonHS,sgdysonR},
as well as for Anderson localization \cite{bovier,molchanov,krit,kuttruf,fyodorov,EBetOG,
fyodorovbis},
on which we focus in this paper.

In the context of Anderson localization \cite{anderson}, the hierarchical models 
\cite{bovier,molchanov,krit,kuttruf,fyodorov,EBetOG,
fyodorovbis} contain long-ranged hoppings
decaying as a power-law of the distance in real space. 
To discuss the possibility of an Anderson localization transition, it is thus useful
to recall first what is known for non-hierarchical long-ranged models.
For Anderson models
with long-ranged hoppings $V(r)$ presenting the typical asymptotic decay
as a function of the distance $r$
\begin{eqnarray}
V(r) \oppropto_{r \to +\infty}  \frac{ V}{r^{\sigma}} 
\label{prbm}
\end{eqnarray}
one expects that Anderson critical points occur in dimension $d$ for
a critical value of the exponent given by 
\begin{eqnarray}
\sigma_c=d
\label{acd}
\end{eqnarray}
whereas $\sigma>d$ correspond to the localized phase (power-law localization)
and $\sigma<d$ corresponds to the delocalized phase
(see the review \cite{mirlinrevue} and references therein).
These critical points have been mostly studied in dimension $d=1$
within the PRBM model (power-law random banded matrices model)
\cite{mirlin96,mirlin_evers,varga00,kra06,garcia06,cuevas01,
cuevas01bis,varga02,cuevas03,mildenberger,mendez05,mendez06,us_transmission},
in particular from the point of view of their multifractal spectra   
(studies in $d>1$ can be found in \cite{potempa,cuevasd,ossipov}).
In hierarchical models, 
one thus also expects the same criticality criterion as in Eq. \ref{acd}.
In the 'ultrametric random matrices ensemble' introduced in \cite{fyodorov},
it has been found that criticality indeed corresponds to $\sigma_c=d=1$ (Eq. \ref{acd})
and multifractality properties of eigenfunctions have been studied \cite{fyodorov,EBetOG,fyodorovbis}.
In particular, in the so-called 'strong multifractality' regime, 
 the powerful Levitov renormalization method \cite{levitov}
(see also the reformulation as
some type of 'virial expansion' in Refs \cite{oleg1,oleg2,oleg3,oleg4})
can be used both for the PRBM model \cite{mirlin_evers} and for the 
'ultrametric random matrices ensemble' \cite{fyodorov} and yield the same leading order result,
which can also be derived via a simpler direct perturbation theory for eigenstates 
\cite{us_strong}.
However the 'ultrametric random matrices ensemble' considered by physicists \cite{fyodorov,EBetOG,fyodorovbis} is not the only possible way to define a Dyson hierarchical model for Anderson localization, and another model containing random on-site energies and non-random hierarchical hoppings, has actually been considered previously by mathematicians \cite{bovier,molchanov,krit,kuttruf}, but in some region of parameters that does not allow to reach the critical point
$\sigma_c=d=1$. As a consequence,
 their results concern the properties of the localized
phase. The aim of the present paper is to show that it is possible to reach the critical point
  $\sigma_c=d=1$ in this type of models, if one chooses properly 
some parameter at a well-defined negative value
(whereas mathematicians \cite{bovier,molchanov,krit,kuttruf} seem to have always considered that this parameter was positive). The difference between the two types of hierarchical models is as follows :
in the model considered in \cite{bovier,molchanov,krit,kuttruf} and in the present paper,
it is directly the matrix of non-random hoppings that presents a hierarchical block structure (see more details in section \ref{sec_model} below),
whereas in References \cite{fyodorov,EBetOG,fyodorovbis}, it is only 
the matrix of the variances of the random hoppings that presents a hierarchical block structure.
Note that in the context of spin models with random couplings, these two 
types of hierarchical models have been also introduced, with either
 a hierarchical structure of the couplings
 \cite{sgdysonAT,sgdysonHS} or a hierarchical structure of the variances of
 the random couplings \cite{sgdysonR}.

The paper is organized as follows.
In section \ref{sec_model}, we introduce the 'Dyson hierarchical model
with random on-site energies and non-random long-ranged hoppings'
that we consider in this paper, and we describe an exact renormalization procedure
in each disordered sample. In section \ref{sec_pure}, we analyse this renormalization
procedure for the pure case, we describe the properties of the energies and eigenstates
as a function of the hierarchical parameter, and explain how to obtain a physical model presenting
a real-space power-law exponent $\sigma_c=1$.
We then turn to the disordered model.
In section \ref{sec_rgonsite}, we analyse
the renormalization of the on-site energies, we find an exact solution for the Cauchy distribution,
and we present numerical results for the box disorder.
In section  \ref{sec_rgtrans}, we describe how the two-point transmission
can be computed via renormalization, and we derive the typical exponent $\kappa_{typ}=2$
independently of the disorder strength. In section 
\ref{sec_falpha}, we translate the previous result into the typical exponent $\alpha_{typ}=2$ for eigenstates, and show numerical results for the singularity spectrum $f(\alpha)$ as a function
of the disorder strength $W$. In section \ref{sec_eigen}, we describe our numerical results
concerning the intermediate statistics of energy levels.
The anomalous weak-disorder regime is discussed in section \ref{weak}.
Our conclusions are summarized in section \ref{sec_conclusion}.

 \section{ Dyson hierarchical model for Anderson localization } 

\label{sec_model}

\subsection{Definition of the model}

\begin{figure}[htbp]
 \includegraphics[height=6cm]{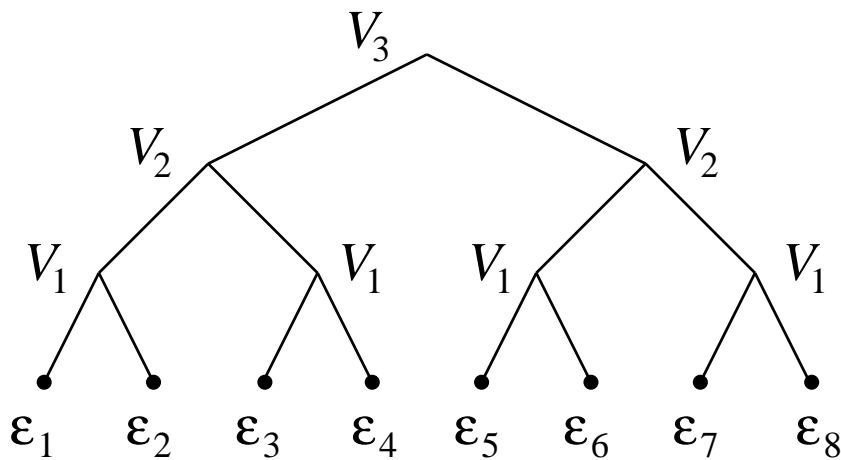}
\caption{ Dyson hierarchical model of Anderson localization defined by the Hamiltonian of Eq. \ref{hamilton} with $N=3$ and $2^N=8$ sites. 
Each site $i$ is characterized by a random on-site energy $\epsilon_i$.
Two sites $i$ and $j$ are related by a non-random hopping $H_{i,j}=V_{n(i,j)}$
where $n(i,j)$ represents the generation $n$ where $i$ and $j$ are related in the binary tree hierarchical structure. For instance, the site $i=1$ is connected to the site $i=2$ via the hopping $V_1$,
to the sites $i=3,4$ via the hopping $V_2$, and to the sites $i=5,6,7,8$ via the hopping $V_3$.
  }
\label{figdysonmodel}
\end{figure}

We consider the following Anderson tight-binding model defined 
 for $L=2^N$ sites by the Hamiltonian (see Fig. \ref{figdysonmodel})
\begin{eqnarray}
H_N [\{\epsilon_1,...,\epsilon_{2^N}\};&& \{V_1,..V_N\}]  =\sum_{i=1}^{2^N} \epsilon_i \vert i > < i \vert  \label{hamilton} \\
  && + V_1 \left[ \left( \vert 1 > < 2 \vert +  \vert 2 > < 1 \vert \right)
+ \left( \vert 3 > < 4 \vert + \vert 4 > < 3 \vert  \right) + ... \right]
\nonumber \\
&& + V_2 \left[  \left( \vert 1 >+\vert 2 > \right)\left(< 3 \vert+< 4 \vert\right) + h.c. + \left( \vert 5 >+\vert 6 > \right)\left(< 7 \vert+< 8 \vert\right) + h.c....
\right] \nonumber \\
&& + V_3 \left[  \left( \vert 1 >+\vert 2 > + \vert 3 >+\vert 4 > \right)\left(< 5 \vert+< 6 \vert +< 7 \vert+< 8 \vert\right)  + h.c. +...
\right] \nonumber \\
&& +V_N \left[  \left( \vert 1 >+\vert 2 > +... + \vert 2^{N-1} > \right)
\left(< 2^{N-1}+1  \vert+< 2^{N-1}+2 \vert +...+< 2^N \vert\right)  + h.c.
\right] \nonumber
\end{eqnarray}
where the $\epsilon_i$ are the independent 
on-site random energies drawn with some distribution,
and where the parameters $(V_1,V_2,...)$ represent the non-random hoppings
at different levels of the hierarchy.
In Dyson hierarchical models, it is usual to take couplings that depend exponentially on the generation $n$
\begin{eqnarray}
V_n = V_1 \gamma^{n-1}
\label{vn} 
\end{eqnarray}
To make the link with the physics of long-ranged one-dimensional models,
it is convenient to consider that the sites $i$ of the Dyson model
 are displayed on a one-dimensional lattice, with a lattice spacing unity.
Then the site $i=1$ is coupled via the hopping $V_n$ to 
the sites $2^{n-1} < i \leq 2^n$. At the scaling level, the hierarchical model
is thus somewhat equivalent to the
following power-law dependence in the real-space distance $L_n=2^n$
\begin{eqnarray}
\vert V_n \vert = \frac{V_1}{\vert \gamma \vert} L_n^{-\sigma(\gamma)} \ \ {\rm with } 
\ \ \sigma (\gamma)=- \frac{\ln \vert \gamma \vert}{\ln 2}
\label{vnabs} 
\end{eqnarray}
It seems physically natural to require here that these hoppings decay with the distance 
$L_n$ with a positive exponent $\sigma(\gamma)>0$ corresponding to $\vert \gamma \vert <1$, i.e. 
\begin{eqnarray}
-1 < \gamma < 1
\label{gammadecaying} 
\end{eqnarray}
We will see later in this paper that other physical requirements may lead to other 
restrictions in the choice of the parameter $\gamma$ (see section \ref{sec_physical}).

\subsection{Exact renormalization procedure}

\label{sec_exactrg}

\subsubsection{ Change of basis }

From the form of the Hamiltonian of Eq. \ref{hamilton},
it is clear that it is useful to begin with a change of basis \\
from the real space natural basis $\{ \vert 1>, \vert 2>, \vert 3>, \vert 4> , ..., \vert 2^N-1 >,\vert 2^N > \}$ \\
to the new basis $\{ \vert 1_+>, \vert 1_->, \vert 2_+>, \vert 2_-> , ..., \vert 2^{N-1}_+ >,\vert 2^{N-1}_- > \}$, 
with the following notation for $1 \leq i \leq 2^{N-1}$
\begin{eqnarray}
\vert i_+ > && \equiv \frac{\vert 2i-1 >+\vert 2i >}{\sqrt 2} \nonumber \\
\vert i_- > && \equiv \frac{\vert 2i-1 >-\vert 2i >}{\sqrt 2} 
\label{pm} 
\end{eqnarray}
In this new basis, the Hamiltonian of Eq. \ref{hamilton} reads
\begin{eqnarray}
H_N [\{\epsilon_1,...,\epsilon_{2^N}\};&& \{V_1,..V_N\}]  =
\sum_{i=1}^{2^{N-1}} \left( \epsilon_{i_+} \vert i_+ > < i_+ \vert
 + \epsilon_{i_-} \vert i_- > < i_- \vert
+ V_{i_{+-}} \left( \vert i_+ > < i_- \vert + \vert i_- > < i_+ \vert  \right) \right) 
\label{hamiltonnewbasis}  \\
&& + 2 V_2  \left[   \vert 1_+ >< 2_+ \vert + \vert 2_+ >< 1_+ \vert +  \vert 3_+> < 4_+ \vert + \vert 4_+> < 3_+ \vert +...
\right] \nonumber \\
&& + 2 V_3 \left[  \left( \vert 1_+ >+\vert 2_+ >  \right)\left(< 3_+ \vert+< 4_+ \vert \right)  + h.c. +...
\right] \nonumber \\
&& +2 V_N \left[  \left( \vert 1_+ >+\vert 2_+ > +... + \vert 2^{N-2}_+ > \right)
\left(< (2^{N-2}+1)_+  \vert+< (2^{N-2}+2)_+ \vert +...+< 2^{N-1}_+ \vert\right)  + h.c.
\right] \nonumber
\end{eqnarray}
in terms of the new parameters
\begin{eqnarray}
\epsilon_{i_+} && \equiv \frac{\epsilon_{2i-1}+\epsilon_{2i} }{2} +V_1 \nonumber \\
\epsilon_{i-}  && \equiv  \frac{\epsilon_{2i-1}+\epsilon_{2i} }{2} -V_1 \nonumber \\
V_{i_{+-}}  && \equiv \frac{\epsilon_{2i-1}-\epsilon_{2i} }{2}
\label{couplingspm} 
\end{eqnarray}

\subsubsection{ Elementary renormalization step }

For Anderson localization models, Aoki \cite{aoki80,aoki82,aokibook} has introduced 
 an exact real-space renormalization procedure at fixed energy which preserves the Green functions of the remaining sites. 
This procedure has been further studied for one-particle models
in \cite{lambert,us_twopoints,us_aokianderson}.
It has been extended in configuration space
for two-particle models \cite{leadbeater} and for  manybody localization models
 \cite{us_manybody}. It can be also used in other physical contexts, like phonons
in random elastic networks \cite{us_phonons}.
Let us now explain how it works for our present purposes.

From the expression of the Hamiltonian in the new basis of Eq. \ref{hamiltonnewbasis},
it is clear that for each $1 \leq i \leq 2^{N-1}$, the vector  $\vert i_- >$
 is only coupled to its symmetric partner $\vert i_+ >$.
As a consequence, if one considers the Schr\"odinger equation at some given energy $E$  
\begin{eqnarray}
E \vert \psi > = H_N \psi >
\label{schro} 
\end{eqnarray}
the projection onto $<i_- \vert$, with the usual notation $<i_{\pm} \vert \psi>=\psi(i_{\pm})$
\begin{eqnarray}
E \psi(i_-) = \epsilon_{i_-} \psi(i_-) +V_{i+-} \psi(i_+)
\label{eliminmoins} 
\end{eqnarray}
can be used to eliminate $\psi(i_-) $ in terms of $\psi(i_+) $.
One obtains for $\vert i_+>$ the renormalized on-site energy
\begin{eqnarray}
\epsilon_{i_+}^{new}(E) && = \epsilon_{i_+}+ \frac{V_{i+-}^2}{E-\epsilon_{i_-}}
\nonumber \\
&& = \frac{\epsilon_{2i-1}+\epsilon_{2i} }{2} +V_1
+ \frac{ \left( \frac{\epsilon_{2i-1}-\epsilon_{2i} }{2}\right)^2}
{E- \left(  \frac{\epsilon_{2i-1}+\epsilon_{2i} }{2} -V_1 \right)}
\label{eplusrg} 
\end{eqnarray}

For the remaining degrees of freedom $\{ \vert 1_+>, \vert 2_+>, \vert 3_+> , ..., \vert 2^{N-1}_+ > \}$, the Hamiltonian reads
\begin{eqnarray}
H_N^{decim} [\{\epsilon_1,...,\epsilon_{2^N}\};\{V_1,..V_N\}] && 
=\sum_{n=1}^{2^{N-1}} \epsilon_{i+}^{new} \vert i+ > < i+ \vert \nonumber \\  
&& + 2 V_2 \left[  \left( \vert 1_+ >< 2_+ \vert + h.c.\right) +
 \left( \vert 3_+ > < 4_+ \vert + h.c \right)....
\right] \nonumber \\
&& + 2 V_3 \left[  \left( \vert 1_+ >+\vert 2_+ >  \right)
\left(< 3_+ \vert+< 4_+ \vert \right)  + h.c. +...
\right] \nonumber \\
&& + 2 V_N \left[  \left( \vert 1_+ >+\vert 2_+ > +... + \vert 2^{N-2}_+ > \right)
\left(< (2^{N-2}+1)_+  \vert +...+< 2^{N-1}_+ \vert\right)  + h.c.
\right]
\nonumber \\
&& = H_{N-1} [\{\epsilon_{1_+}^{new},...,\epsilon_{2^{N-1}_+}^{new} \};
 \{V_1'=2 V_2,..,V_{N-1}'=2 V_N\}]  
\label{hamilton1pas} 
\end{eqnarray}

In summary, the Hamiltonian defined for the $2^N$ initial sites with the random 
 energies $(\epsilon_1,\epsilon_2,...,\epsilon_{2^N})$ and the hierarchical hoppings
$(V_1,V_2,...V_N)$ leads, after the decimation of the $2^{N-1}$ antisymmetric 
combinations $\{ \vert 1_->, \vert 2_->, ..., \vert 2^{N-1}_-> \}$, to the same form of Hamiltonian for the remaining $2^{N-1}$ symmetric 
combinations $\{ \vert 1_+>, \vert 2_+>, ..., \vert 2^{N-1}_+> \}$,
with renormalized parameters : the renormalized energies are given by the rule of Eq. \ref{eplusrg}, whereas the hierarchical couplings are simply given by 
$V_1'=2 V_2, ..,V_{N-1}'=2 V_N$.

Before we consider the disordered case, let us first describe in the next section 
the properties of the pure model.

\section{ Properties of the pure model $\epsilon_i=\epsilon_0$ }

\label{sec_pure}

In this section, we describe the properties of the pure model, where all on-site energies
in the Hamiltonian of Eq. \ref{hamilton} take the same value $\epsilon_i=\epsilon_0$,
and where the hierarchical couplings follow the geometric form of Eq. \ref{vn}.

\subsection{ Enumeration of eigenstates  }

\label{enumeration}

We follow the renormalization procedure introduced in the previous section \ref{sec_exactrg}.
In the pure case, the parameters after the first change of basis do not depend on $i$ and
simply read (Eq. \ref{couplingspm}) 
\begin{eqnarray}
\epsilon^{(1)}_{+} && = \epsilon_0 +V_1 \nonumber \\
\epsilon^{(1)}_{-}  && = \epsilon_0  -V_1 \nonumber \\
V_{{+-}}^{(1)}  && = 0
\label{couplingspmpure1} 
\end{eqnarray}
Since the intercoupling vanishes $V_{{+-}}^{(1)}=0 $, the $2^{N-1}$ states $\vert i_- >$
are completely decoupled : they are exact eigenstates of energies $\epsilon^{(1)}_{-}$.
So in the pure case, the decimation step is completely trivial, and there is no renormalization of the on-site energies for the remaining states $\vert i_+>$, 
( Eq. \ref{eplusrg} reduces to the identity $ \epsilon_{i_+}^{new}(E) = \epsilon_{i_+}$). 

The second renormalization step is thus similar, but with the new values
(using $V_1'=2V_2$ from Eq. \ref{hamilton1pas})
\begin{eqnarray}
\epsilon^{(2)}_{+} && = \epsilon^{(1)}_{+} +V_1'= \epsilon_0 +V_1 +2 V_2 \nonumber \\
\epsilon^{(2)}_{-}  && = \epsilon^{(1)}_{+}  -V_1' = \epsilon_0 +V_1 -2 V_2
\label{couplingspmpure2} 
\end{eqnarray}
So by iteration, the n-th renormalization step is again similar, but with the new values
\begin{eqnarray}
\epsilon^{(n)}_{+} && = \epsilon^{(n-1)}_{+} + 2^{n-1}V_n= \epsilon_0 +V_1 +2 V_2+...+2^{n-2} V_{n-1}+ 2^{n-1}V_n \nonumber \\
\epsilon^{(n)}_{-}  && = \epsilon^{(n-1)}_{+} - 2^{n-1}V_n = \epsilon_0 +V_1  +2 V_2+... +2^{n-2} V_{n-1} - 2^{n-1}V_n
\label{couplingspmpuren} 
\end{eqnarray}
When the hierarchical couplings $V_n$ follow the geometric form of Eq. \ref{vn},
these energies simply read 
\begin{eqnarray}
\epsilon_+^{(n)} && =\epsilon_0+V_1 \sum_{i=1}^n (2 \gamma)^{i-1}=\epsilon_0+V_1 \frac{1-(2\gamma)^n}{1-2\gamma}
\nonumber \\
\epsilon_-^{(n)} && =\epsilon_+^{(n-1)}-2^{n-1} V_n = \epsilon_0+V_1 \frac{1 - (2\gamma)^{n-1} 2 (1-\gamma)}{1-2\gamma}
\label{epspmpur} 
\end{eqnarray}

We may now enumerate all the eigenstates, or more precisely all the subspaces 
associated to eigenvalues.

\subsubsection{ Subspace of dimension $2^{N-1}$ associated with the energy $\epsilon_-^{(1)}$  }

We have seen above that the subspace associated to the energy $\epsilon_-^{(1)}$
can be constructed from the basis (Eq \ref{pm})
\begin{eqnarray}
\vert i_- >^{(1)} = \frac{\vert 2i-1 >-\vert 2i >}{\sqrt 2} 
\label{1moins} 
\end{eqnarray}
with $i=1,2,...,2^{N-1}$.
Each vector $\vert i_- >^{(1)} $ is thus extremely 'localized' : its support
is made of two sites only.
This property may seem very strange by comparison with non-hierarchical
translation invariant models, where eigenstates are given by plane-waves that are 
completely delocalized over the whole lattice. 
However here also we can make plane-wave linear combinations of the basis of Eq. \ref{1moins}
to obtain a new-basis (noting $j=\sqrt{-1}$)
\begin{eqnarray}
\vert q >^{(1)} = \frac{1}{\sqrt{2^{N-1}}} \sum_{i=1}^{2^{N-1}} e^{j q i} \vert i_- >^{(1)}
\label{1q}  
\end{eqnarray}
with the wave-vector $q= m (2 \pi/2^{N-1} )$ where $m$ takes the integer values $m=0,1,..,2^{N-1}-1$. Each vector $\vert q >^{(1)} $ is then delocalized on the whole lattice, with a constant weight $\vert < i \vert q >^{(1)} \vert^2=1/2^N$.

\subsubsection{ Subspace of dimension $2^{N-2}$ associated with the energy $\epsilon_-^{(2)}$  }

 The subspace associated to the energy $\epsilon_-^{(2)}$
can be constructed from the basis 
\begin{eqnarray}
\vert i_- >^{(2)} = \frac{\vert (2i-1)_+ >^{(1)} -\vert (2i)_+ >^{(1)} }{\sqrt 2} 
= \frac{\vert 4 i-3 > + \vert 4 i-2 > -\vert 4 i-1 > -\vert 4 i > }{ 2} 
\label{2moins} 
\end{eqnarray}
with $i=1,2,...,2^{N-2}$.  
Each vector $\vert i_- > $ is 'localized' on four sites only.
But again, one may use instead a basis of delocalized plane-waves
\begin{eqnarray}
\vert q >^{(2)} = \frac{1}{\sqrt{2^{N-2}}} \sum_{i=1}^{2^{N-2}} e^{j q i} \vert i_- >^{(2)}
\label{2q} 
\end{eqnarray}
with the wave-vector $q= m (2 \pi/2^{N-2} )$ where $m$ takes the integer values
 $m=0,1,..,2^{N-2}-1$. 

It is now clear how this construction can be pursued, and we only mention the 
last two generations.

\subsubsection{ Subspace of dimension $2$ associated with the energy $\epsilon_-^{(N-1)}$  }

 The subspace associated to the energy $\epsilon_-^{(N-1)}$
can be constructed from the basis 
\begin{eqnarray}
\vert 1_- >^{(N-1)} = \frac{\vert 1_+ >^{(N-2)} -\vert 2_+ >^{(N-2)} }{\sqrt 2} 
= \frac{ 1}{ \sqrt{2^{N-1} }} \left( \sum_{i=1}^{2^{N-2}} \vert i > - \sum_{i=2^{N-2}+1}^{2^{N-1}}\vert i > \right)
  \nonumber \\
\vert 2_- >^{(N-1)} = \frac{\vert 3_+ >^{(N-2)} -\vert 4_+ >^{(N-2)} }{\sqrt 2} 
= \frac{1 }{ \sqrt{2^{N-1} }} \left(  \sum_{i=2^{N-1}+1}^{2^{N-1}+2^{N-2}} \vert i > - \sum_{i=2^{N-1}+2^{N-2}+1}^{2^{N}}\vert i > \right)
\label{beforelastmoins} 
\end{eqnarray}

\subsubsection{ Subspaces of dimension $1$ associated with $\epsilon_-^{(N)}$ and $\epsilon_+^{(N)}$ }

The last renormalization step yields that the energies $\epsilon_-^{(N)}$ and $\epsilon_+^{(N)}$
are non-degenerate and are associated respectively with the eigenstates
\begin{eqnarray}
\vert 1_- >^{(N)} = \frac{\vert 1_+ >^{(N-1)} -\vert 2_+ >^{(N-1)} }{\sqrt 2} 
= \frac{1  }{ \sqrt{2^{N} }} \left( \sum_{i=1}^{2^{N-1}} \vert i > - \sum_{i=2^{N-1}+1}^{2^{N}}\vert i > \right)
\label{lastmoins} 
\end{eqnarray}
and
\begin{eqnarray}
\vert 1_+ >^{(N)} = \frac{\vert 1_+ >^{(N-1)} +\vert 2_+ >^{(N-1)} }{\sqrt 2} 
= \frac{1  }{ \sqrt{2^{N} }} \sum_{i=1}^{2^{N}} \vert i >
\label{lastplus} 
\end{eqnarray}

One may check that the total number of states is $2^N$ as it should
\begin{eqnarray}
\left[ 2^{N-1}+2^{N-2}+...+2 +1 \right] +1 = \frac{2^N-1}{2-1} +1=2^N
\label{comptage} 
\end{eqnarray}

\subsection{ Green function of the pure case }

From this enumeration of eigenstates, one obtains directly
that the Green function at coinciding points reads
\begin{eqnarray}
G_{1,1}(z) = \sum_{\alpha} \frac{ \vert \psi_{\alpha}(1) \vert^2 }{z-E_{\alpha}}
= \sum_{n=1}^N \frac{ \frac{1}{2^n} }{z-\epsilon_-^{(n)}} 
+ \frac{ \frac{1}{2^N} }{z-\epsilon_+^{(N)}}
\label{Greencoincidingpure} 
\end{eqnarray}
(because at the given point $1$, only one state of each generation 
gives a non-zero weight $\vert \psi_{\alpha}(1) \vert^2$).

More generally, the Green function between two distinct points $(1,2^p)$
has a similar form as Eq. \ref{Greencoincidingpure}, except that the sum begins
only with the state of generation $p$ which is non-vanishing both at  $1$ and $2^p$
\begin{eqnarray}
G_{1,2^p}(z) = \sum_{\alpha} \frac{  \psi_{\alpha}^*(1) \psi_{\alpha}^*(2^p) }{z-E_{\alpha}}
= \sum_{n=p}^N \frac{ \frac{1}{2^n} }{z-\epsilon_-^{(n)}} 
+ \frac{ \frac{1}{2^N} }{z-\epsilon_+^{(N)}}
\label{Greenpure} 
\end{eqnarray}

\subsection{ Density of states of the pure case }

\label{secdospure}

From the Green function at coinciding points of Eq. \ref{Greencoincidingpure},
one obtains that the density of states in the thermodynamic limit $N \to +\infty$ reads
\begin{eqnarray}
\rho(E) = \sum_{n=1}^{+\infty}  \frac{1}{2^n} \delta \left( E - \epsilon_-^{(n)} \right)
\label{dospure} 
\end{eqnarray}
(the normalization is $\int dE \rho(E) = \sum_{n=1}^{+\infty}  \frac{1}{2^n} =1$ as it should).
So in contrast to usual non-hierarchical models in finite dimensions 
that are characterized by a continuum of non-degenerate delocalized states, 
the density of states of the hierarchical model in the thermodynamic limit 
 remains a sum of delta functions, with highly degenerate energies.

Since the levels $\epsilon_-^{(n)} $ are given by Eq. \ref{epspmpur}
\begin{eqnarray}
\epsilon_-^{(n)}  && = \epsilon_0 +V_1 \left[ 1  +2\gamma +... +(2 \gamma)^{n-2} 
 - (2 \gamma)^{n-1} \right] \nonumber \\
&& =  \epsilon_0+V_1 \frac{1 - (2\gamma)^{n-1} 2 (1-\gamma)}{1-2\gamma}
\label{emoinspur} 
\end{eqnarray}
various behaviors are possible as a function of the hierarchical parameter $\gamma$
introduced in Eq. \ref{vn} (recall the condition $ \vert \gamma \vert <1 $ of Eq. \ref{gammadecaying} ):

(i) for $  1/2 < \vert \gamma \vert < 1$ : the absolute values 
of the energies $ \vert \epsilon_-^{(n)} \vert$ grow exponentially in $n$.
This case is thus unphysical, since higher levels of the hierarchy that corresponds to
more extended states are associated with higher and higher energies.

(ii) for $ \vert \gamma \vert < 1/2$  : as $n$ grows, the energies $ \epsilon_-^{(n)}$ accumulate near the finite accumulation point 
\begin{eqnarray}
\epsilon_-^{acc}  = \epsilon_0+V_1 \frac{1}{1-2\gamma}
\label{emoinspuracc} 
\end{eqnarray}

(iii) for the special case $\gamma=1/2 $ : the absolute values 
of the energies $ \vert \epsilon_-^{(n)} \vert$ grow linearly in $n$
\begin{eqnarray}
\epsilon_-^{(n)}   = \epsilon_0 +V_1 \left[ n-2 \right]
\label{emoinspurdemi} 
\end{eqnarray}
This case is thus unphysical as the case (i) discussed above.

(iv) for the special case $\gamma=-1/2 $ : the energies $ \epsilon_-^{(n)}$ take alternatively two values,
and the model is thus 'physical' in contrast to (iii).

\subsection{ Physical region of the model }

\label{sec_physical}

From the above discussion, the requirement to have a bounded density of states
for the discrete model of Eq. \ref{hamilton} with the hierarchical couplings of Eq. \ref{vn} 
yields the following domain for the hierarchical parameter $\gamma$
\begin{eqnarray}
- \frac{1}{2} \leq \gamma < \frac{1}{2}
\label{domainegammademi} 
\end{eqnarray}
instead of Eq. \ref{gammadecaying}.

From the point of view of the real-space power-law exponent $\sigma(\gamma)$ of Eq. \ref{vnabs},
our conclusion is thus that the domain $\sigma(\gamma)<1$ corresponding to $  1/2 < \vert \gamma \vert < 1$
(case (i) above) is not physical, whereas the domain $\sigma(\gamma)>1$ corresponding to $ \vert \gamma \vert < 1/2$ (case (ii) above) is physical.
Finally the critical value $\sigma_c=1$ of Eq. \ref{acd}
cannot be made physical with the choice $\gamma=1/2$ (case (iii) above),
but can be made physical with the choice $\gamma=-1/2$ (case (iv) above).

In the presence of disorder, the criterion of criticality for Anderson localization
in the presence of long-ranged hoppings that we have recalled in the introduction around Eq. \ref{acd}
leads to the following conclusion : 
all cases with $- \frac{1}{2} < \gamma < \frac{1}{2}$ are expected to be in the localized phase,
whereas the case 
\begin{eqnarray}
\gamma_c = - \frac{1}{2} 
\label{gammac} 
\end{eqnarray}
is expected to be critical.
In the remaining of this paper, we thus focus on this case,
which has not been considered previously in
the mathematical literature as we now explain.

\subsection{ Relations with the models considered in the mathematical literature }

\subsubsection{ Relation with the model of Ref \cite{bovier} }

Eq. \ref{Greencoincidingpure} is in agreement with formula  (2.25) of Bovier \cite{bovier}
in the thermodynamic limit $N \to +\infty$, with a simple shift $n \to n+1$
in the labelling of $\epsilon_-^{(n)}$ given in Eq. \ref{epspmpur},
with the following correspondence of notations : Bovier's model corresponds to the parameters
 $\epsilon_0=0$, $V_1=-1/2$ and
\begin{eqnarray}
\alpha=4 \gamma
\label{epsbovierpur} 
\end{eqnarray}
Bovier has assumed $\alpha>0$, in order to interpret the pure Hamiltonian $H_0=-\Delta_{\alpha}$
as a 'hierarchical Laplacian'.
However if one defines the model via a tight-binding quantum Hamiltonian as we did in Eq. \ref{hamilton}, there is no reason from a physical point of view
to impose a priori some sign constraints on the hoppings $V_n$, i.e. on $\alpha=4 \gamma$.
So the interesting critical case of Eq. \ref{gammac} corresponds to $\alpha=-2$ in the notations of
\cite{bovier}.

\subsubsection{ Relation with the model of Refs \cite{molchanov,krit,kuttruf} }

In Refs \cite{molchanov,krit,kuttruf}, the notion of 'hierarchical Laplacian' has been defined
in terms of a sequence of positive numbers $p_n \geq 0$ for $n=1,2,..$ such that 
\begin{eqnarray}
\sum_{n=1}^{+\infty} p_n=1
\label{normapn} 
\end{eqnarray}
Translated into our present notations with the quantum Hamiltonian of Eq. \ref{hamilton},
this version of the model corresponds to the pure on-site energy
\begin{eqnarray}
\epsilon_0= \sum_{n=1}^{+\infty} \frac{p_n}{2^n}
\label{epszeromath} 
\end{eqnarray}
and to the hierarchical hoppings
\begin{eqnarray}
V_n= \sum_{m=n}^{+\infty} \frac{p_m}{2^m}
\label{vnmath} 
\end{eqnarray}
so that the eigenenergies of Eq \ref{couplingspmpuren} reads
\begin{eqnarray}
\epsilon^{(n=1)}_{-} && = \epsilon_0 - V_1 =0 \nonumber \\
\epsilon^{(n=2)}_{-} &&= \epsilon_0 + V_1 -2 V_2 = p_1 \nonumber \\
\epsilon^{(n=3)}_{-} && = \epsilon_0 + V_1 +2 V_2 - 4 V_3= p_1+p_2 \nonumber \\
\epsilon^{(n)}_{-}  && =  \epsilon_0 +V_1  +2 V_2+... +2^{n-2} V_{n-1} - 2^{n-1}V_n = \sum_{m=1}^{n-1} p_m
\label{eigenmath} 
\end{eqnarray}
where the accumulation point is fixed to (Eq. \ref{normapn})
\begin{eqnarray}
\epsilon^{acc}=\sum_{n=1}^{+\infty} p_n=1
\label{eaccmath} 
\end{eqnarray}
 Here again, Refs \cite{molchanov,krit,kuttruf} impose the positivity of the numbers $p_n \geq 0$
to have an interpretation in terms of some classical random walk for the hierarchical Laplacian,
whereas if one defines the model via a tight-binding quantum Hamiltonian as we did in Eq. \ref{hamilton}, there is no reason to impose the positivity of all the $p_n$, and the only physical constraint
is on the eigenvalues $\epsilon^{(n)}_{-} $ that should remain bounded, i.e. 
the constraint is that all partial sums $\sum_{m=1}^{n-1} p_m $ should remain bounded.
In particular, the interesting critical case of Eq. \ref{gammac} corresponds
to the choice $p_n=(-1)^{n-1}$.

\subsection{ Special case $\gamma_c=-1/2 $ corresponding to a real-space power-law of exponent $\sigma_c=1$}

\label{sec_purecriti}

For the special case $\gamma_c=-1/2$, there exist an exact invariance after two steps of RG,
and Eq \ref{epspmpur} become
\begin{eqnarray}
\epsilon_+^{(n)} && =\epsilon_0+V_1 \frac{1-(-1)^n}{2}
\nonumber \\
\epsilon_-^{(n)} && = \epsilon_0+V_1 \frac{1 - (-1)^{n-1} 3 }{2}
\label{epspmpurcriti} 
\end{eqnarray}
As a consequence, in the enumeration of eigenstates of section \ref{enumeration},
the states $(-)$ have only two possible energies, corresponding to odd and even generations
\begin{eqnarray}
\epsilon_-^{odd}&& = \epsilon_0- V_1
\nonumber \\
\epsilon_-^{even} && = \epsilon_0+ 2 V_1
\label{emdegecriti} 
\end{eqnarray}
whereas the last state $(+)$ has for energy
\begin{eqnarray}
\epsilon_+^{(N odd)} && =\epsilon_0+V_1 
\nonumber \\
\epsilon_+^{(N even)} && =\epsilon_0
\label{epcritimaxi} 
\end{eqnarray}

In the thermodynamic limit $N \to +\infty$, the Green function of Eq \ref{Greenpure} becomes
\begin{eqnarray}
G_{1,2^p}(z)
= \sum_{n=p}^{+\infty} \frac{ \frac{1}{2^n} }{z-\epsilon_-^{(n)}} 
=  \frac{ g^{odd}(2^p) }{z-\epsilon_-^{odd}} 
+ \frac{ g^{even}(2^p) }{z-\epsilon_-^{even}} 
\label{Greenpurecriti} 
\end{eqnarray}
with the coefficients 
\begin{eqnarray}
g^{odd}(2^p) && = \sum_{k=0}^{+\infty}  \frac{\theta(2k+1 \geq p)}{2^{2k+1}}  \nonumber \\
g^{even}(2^p) && = \sum_{k=2}^{+\infty}  \frac{\theta(2k \geq p)}{2^{2k}}  
\label{goddeven} 
\end{eqnarray}
that decay as $1/2^p=1/L_p$ with respect to the distance $L_p=2^p$.

In particular at coinciding points corresponding to $p=0$ ($2^p=1$), one obtains
the simple values
\begin{eqnarray}
g^{odd}(1) && = \sum_{k=0}^{+\infty}  \frac{1}{2^{2k+1}} = \frac{2}{3} \nonumber \\
g^{even}(1) && = \sum_{k=2}^{+\infty}  \frac{1}{2^{2k}} = \frac{1}{3}
\label{goddevenpzero} 
\end{eqnarray}
so that the density of states of Eq. \ref{dospure}
is simply given by the following sum of two delta peaks
\begin{eqnarray}
\rho_{pure}(E) =  \frac{2}{3} \delta \left( E - \epsilon_-^{odd} \right) + \frac{1}{3} \delta \left( E - \epsilon_-^{even} \right)
\label{dospurecriti} 
\end{eqnarray}
This pure model is thus extremely degenerate.
As a consequence, the effect of a 'weak disorder' is expected to be anomalous at the perturbative level (see more details in section \ref{weak}).

\section{ Analysis of the renormalization of the on-site energies }

\label{sec_rgonsite}

In this section, we analyse the renormalization procedure of section \ref{sec_exactrg}
when the initial on-site energies are random variables.
We first consider the case of the Cauchy distribution that leads to an explicit exact solution.
We then study numerically the renormalization flow starting from a box distribution.

\subsection{ Cauchy disorder : exact solution }

\subsubsection{ Reminder on some specific properties of the Cauchy distribution }

\label{cauchyprop}

As is well-known, the Cauchy distribution is the only probability distribution which is stable both by addition and by inversion.
More precisely, if one denotes the Cauchy distribution
of mean $a$ and width $b$ by
\begin{eqnarray}
C_{a,b}(x) \equiv \frac{1}{\pi b } \ \frac{1}{1+ \left( \frac{x-a}{b} \right)^2}
\label{cauchy} 
\end{eqnarray}
one has the two properties :

(i) addition : if $x_1$ is distributed with $C_{a_1,b_1}(x_1)$ 
and $x_2$ is distributed with $C_{a_2,b_2}(x_2)$, then $x=x_1+x_2$
is distributed with $C_{a,b}(x)$ of parameters $a=a_1+a_2$ and $b=b_1+b_2$.

(ii) inversion : if $x$
is distributed with $C_{a,b}(x)$, then $y=1/x$ is distributed with 
  $C_{A,B}(y)$ of parameters $A=a/(a^2+b^2)$ and $B=b/(a^2+b^2)$.

\subsubsection{ Renormalization flow of the on-site energies $\epsilon_{i\pm} $ }

The RG Equation \ref{eplusrg}, that describes how the on-site energies evolve upon 
the first RG step
(Eq. \ref{hamilton1pas}), can be rewritten as
\begin{eqnarray}
\frac{1}{ \frac{(\epsilon_{i+}^{new}-E)}{2}-V_1}  =
\frac{1}{ (\epsilon_{2i-1}-E )-V_1}
+\frac{1}{ (\epsilon_{2i}-E )-V_1}
\label{eplusrgsimplified} 
\end{eqnarray}
Let us recall the notations used in this relation : 
the energy $E$ and and the hopping $V_1$ are parameters, 
$\epsilon_{2i-1}$ and $\epsilon_{2i}$ are independent on-site energies
 of the initial model, and 
$\epsilon_{i+}$ is a renormalized on-site energy 
after one renormalization step.

We now apply the properties of the Cauchy distribution recalled above in section \ref{cauchyprop}.
We assume that the two independent on-site energies
$\epsilon_{2i-1}$ and $\epsilon_{2i}$ of the first generation 
are drawn with the distribution $C_{a_1,b_1}(\epsilon)$ with the values
\begin{eqnarray}
a_1=\epsilon_0 \nonumber \\
b_1= W
\label{a1b1} 
\end{eqnarray}
where $W$ represents the disorder strength around the averaged value $\epsilon_0$.

As a consequence, one obtains that

(1) $(\epsilon_{2i-1}-E-V_1)$ and $(\epsilon_{2i}-E-V_1)$
are distributed with $C_{a_1-E-V_1,b_1}$

(2) $1/(\epsilon_{2i-1}-E-V_1)$ and $1/(\epsilon_{2i}-E-V_1)$
are distributed with $C_{A,B}$ of parameters
\begin{eqnarray}
A && =\frac{a_1-E-V_1}{(a_1-E-V_1)^2+b_1^2} \nonumber \\
B && =\frac{b_1}{(a_1-E-V_1)^2+b_1^2}
\label{AB} 
\end{eqnarray}

(3) the sum $S=1/(\epsilon_{2i-1}-E-V_1)+1/(\epsilon_{2i}-E-V_1)$
is distributed with $C_{2A,2B}$.

(4) the inverse $I=1/S$ is distributed with $C_{a_I,b_I}$
of parameters
\begin{eqnarray}
a_I && =\frac{2A}{(2A)^2+(2B)^2} = \frac{a_1-E-V_1}{2} \nonumber \\
b_I && =\frac{2B}{(2A)^2+(2B)^2} = \frac{b_1}{2}
\label{ABI} 
\end{eqnarray}

(5) the on-site energy $\epsilon_{i+}^{new}=E +2V_1 +2 I$ after one RG step
 is then distributed with $C_{a_2,b_2}$
of parameters
\begin{eqnarray}
a_2 && = E+2 V_1+2 a_I = a_1+V_1   \nonumber \\
b_2 && = 2 b_I = b_1
\label{AB2} 
\end{eqnarray}
i.e. the width of the distribution is conserved, whereas the averaged value in shifted by $V_1$.

Taking into account the evolution of the lowest coupling
after $(n-1)$ renormalization steps (see Eq. \ref{hamilton1pas})
 we obtain that if the on-site energies of the initial model 
are distributed with the Cauchy distribution $C_{a_1,b_1}$ with Eq. \ref{a1b1},
 then the on-site energies
after $n$ renormalization steps are
distributed with the Cauchy distribution $C_{a_n,b_n}$
of parameters
\begin{eqnarray}
a_{n+1} && = a_n+V_1^{(n-1)} =\epsilon_0+V_1^{(0)}+ V_1^{(1)}... +V_1^{(n-1)}
=a_1+ V_1+2 V_2 + 2^2 V_3 +...+ 2^{n-1} V_n
   \nonumber \\
b_{n+1} && =  b_n = b_1 = W
\label{anbn} 
\end{eqnarray}

For $W=0$, we recover of course the pure model analyzed previously in section \ref{sec_pure}.
In particular, the physical cases where the on-site energies do not flow to infinity 
correspond to the domain of Eq. \ref{domainegammademi}.

\subsubsection{ Critical case $\gamma_c=-1/2$ : exact invariance after two RG steps }

As previously discussed in section \ref{sec_purecriti} concerning the pure case,
the special case $\gamma_c=-1/2$ corresponding to a real-space power-law of exponent $\sigma_c=1$
has the special property to be exactly invariant after two RG steps.
This property survives in the disordered case
for the Cauchy distribution : 
 the parameters of the Cauchy distribution of Eq. \ref{anbn} 
are stable after two renormalization steps
\begin{eqnarray}
a_{2n+1} && \equiv a_{odd} = a_1 = \epsilon_0 \nonumber  \\
a_{2n} && \equiv a_{even} = a_1+V_1 = \epsilon_0 +V_1
\label{anoddeven} 
\end{eqnarray}

\subsection{ Case of other initial on-site energy distributions at criticality $\gamma_c=-1/2 $ }

\begin{figure}[htbp]
 \includegraphics[height=6cm]{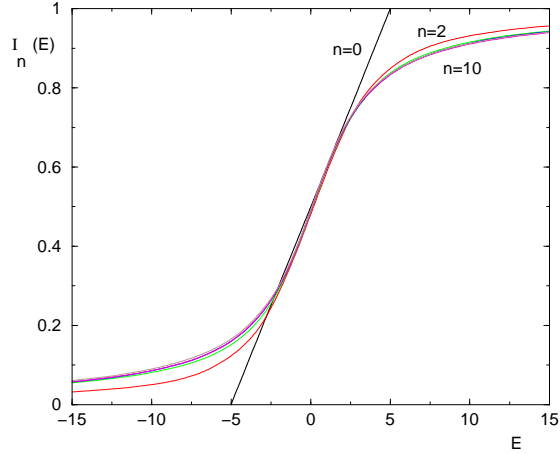}
\caption{ Integrated distribution $I_n(E) \equiv 
\int_{-\infty}^E d\epsilon P_n(\epsilon)$
of the probability distribution $P_n(\epsilon)$ of
the renormalized on-site energies obtained via the RG rule of Eq.
\ref{ergcriti} as a function of the generation $n=2,4,6,8,10$ starting at $n=0$ from the box distribution of Eq. \ref{box} with $W=5$ : the convergence towards the integrated Cauchy distribution $ I_{Cauchy}(E)=\int_{-\infty}^E dx \frac{b}{\pi \left[ (x-a)^2+b^2 \right]} = \frac{1}{2}+  \frac{1}{\pi} \arctan \left( \frac{E-a}{b}\right)$ is very fast.
  }
\label{figrgeps}
\end{figure}

As explained in the introduction around Eq. \ref{acd}, one expects that a model
with real-space power-law hopping $\sigma=1$ will be critical for any on-site energy distribution.
In our present case where $\sigma_c=1$ is realized for the value $ \gamma_c=-1/2$,
we thus expect criticality to occur for any initial distribution of the on-site energies.
It is then interesting to study the renormalization flow of the on-site energies.
As an example, we have considered the case where the initial on-site energies are drawn 
with the box distribution
\begin{eqnarray}
P_W^{Box}(\epsilon) = \frac{1}{W} \theta \left( - \frac{W}{2} \leq \epsilon \leq \frac{W}{2} \right)
\label{box} 
\end{eqnarray}
and we have computed numerically the probability distribution of renormalized on-site energies
obtained by the RG equation for $\gamma_c=-1/2$ 
\begin{eqnarray}
\frac{1}{ \frac{(\epsilon_{i}^{(n)}-E)}{2}- V_1(-1)^{n-1}}  =
\frac{1}{ (\epsilon_{2i-1}^{(n-1)}-E )-V_1(-1)^{n-1}}
+\frac{1}{ (\epsilon_{2i}^{(n-1)}-E )-V_1(-1)^{n-1}}
\label{ergcriti} 
\end{eqnarray}
As an example, we show on Fig. \ref{figrgeps}
the numerical results corresponding $V_1=1$, $E=0$ starting from
the initial distribution of Eq. \ref{box} with $W=10$ : 
after a small number $n=2,4,6,8,10$ of RG steps, 
we obtain a very rapid convergence towards a Cauchy-like distribution. 
Our conclusion is thus that the Cauchy distributions found above seem to be the only attractive fixed
points.

 \section{ Renormalization procedure to study the transmission properties }

\label{sec_rgtrans}

\begin{figure}[htbp]
 \includegraphics[height=5cm]{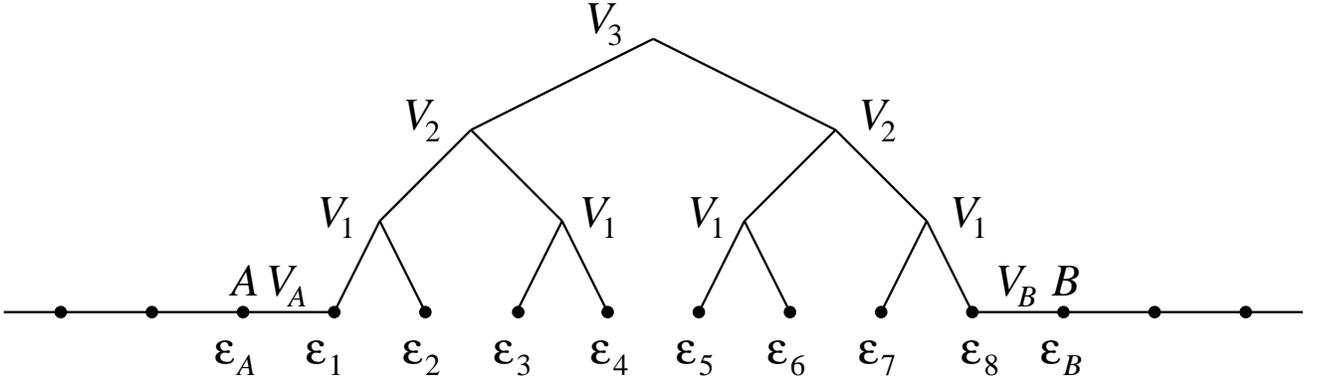}
\caption{ The model of Fig. \ref{figdysonmodel} is now connected to an incoming wire ending at point $A$
connected to the site $i=1$, and to an outgoing wire ending at point $B$ connected to the site $i=2^N$ with $N=3$ here.
  }
\label{figdysontrans}
\end{figure}

In quantum coherent problems, a convenient characterization 
of transport properties consists in defining a scattering problem
where the disordered sample is linked to incoming wires and outgoing wires,
and in studying the reflection and transmission coefficients.
This scattering theory definition of transport, 
first introduced by Landauer \cite{landauer},
has been often used for one-dimensional systems 
\cite{anderson_fisher,anderson_lee,luck}
and has been generalized to higher dimensionalities and multi-probe
measurements (see the review \cite{stone}).

In this section, we assume that the first site $i=1$ is connected to an incoming wire,
and that the last site $i=2^N$ is connected to an outgoing wire (see Figure \ref{figdysontrans}).

 \subsection{ Elementary RG step for the couplings to the external wires }

We call $A$ the point of the wire of on-site energy $\epsilon_A$
that is connected to the site $i=1$ via the hopping $V_{A,1}$.
During the change of basis of Eq. \ref{pm}, the hopping between $A$ and $1$ 
becomes
\begin{eqnarray}
 V_{A} \left( \vert A> < 1 \vert + \vert 1> < A \vert \right)
= 
V_{A+} \left( \vert A> < 1_+ \vert + \vert 1_+> < A \vert \right)
 +  V_{A-} \left(\vert A> < 1_- \vert
+ \vert 1_-> < A \vert \right)
\label{apm} 
\end{eqnarray}
with
\begin{eqnarray}
V_{A+} =V_{A-} = \frac{V_{A}}{\sqrt{2}} 
\label{couplingsApm} 
\end{eqnarray}
Since  $\vert 1_- >$ is only coupled to $\vert 1_+ >$ and $\vert A >$,
the Schr\"odinger equation at energy $E$ 
\begin{eqnarray}
E \psi(1_-) = \epsilon_{1_-} \psi(1_-) +V_{1+-} \psi(1_+)+V_{A-} \psi(A)
\label{eliminmoinsbis} 
\end{eqnarray}
can be used to eliminate $\psi(i_-)$ in terms of $\psi(i_+)$ and $\psi(A)$, 
so that in addition to Eqs \ref{eplusrg}, the renormalized on-site energy of site $A$
and the renormalized coupling $V_{A+}^{new}$
read using Eqs \ref{couplingspm}
\begin{eqnarray}
\epsilon_{A}^{new} && = \epsilon_{A}+ \frac{V_{A-}^2}{E-\epsilon_{1_-}}
= \epsilon_{A} + \frac{V_A^2}{2 (E+V_1) - (\epsilon_{1}+\epsilon_{2} )}
\nonumber \\
V_{A+}^{new} && = V_{A+} + \frac{V_{A-}V_{-+}}{E-\epsilon_{1_-}}
= \frac{V_A}{\sqrt 2} \left[ 1+ \frac{\epsilon_{1}-\epsilon_{2}}{2(E+V_1)-(\epsilon_{1}+\epsilon_{2} )} \right]
\label{rgrulesAtrans}
\end{eqnarray}
In the following, the important point is that the 
the renormalization rule  for the renormalized coupling to A can be rewritten
as
\begin{eqnarray}
\frac{ V_{A+}^{new} }{ V_A } 
= \frac{1}{\sqrt 2}
 \left[ \frac{\epsilon_2-E-V_1}
{ \frac{\epsilon_{1}+\epsilon_{2}}{2}-E-V_1} \right] 
\label{rgVAtrans}
\end{eqnarray}

Similarly, the hopping $V_B$ between $B$ and the last site $2^N$
is renormalized during the first RG step into
\begin{eqnarray}
\frac{ V_{B+}^{new} }{ V_B } 
= \frac{1}{\sqrt 2}
 \left[ \frac{\epsilon_{(2^N-1)}-E-V_1}
{ \frac{\epsilon_{(2^N-1)}+\epsilon_{2^N}}{2}-E-V_1} \right] 
\label{rgVBtrans}
\end{eqnarray}

 \subsection{ Effective transmission for a finite system of size $L_N=2^N$}

We consider a finite system of size $L_N=2^N$.
During the first $(N-1)$ RG steps, the hoppings $V_A$ and $V_B$ to the external wires
are renormalized independently. 
Denoting by $\epsilon_{1,2}^{(n)}$ the first two renormalized on-site energies after $(n-1)$ RG steps
(so that the initial on-site energies correspond to $ \epsilon_{1}^{(n=1)}=\epsilon_1$
and $ \epsilon_{2}^{(n=1)}=\epsilon_2$),
 and $V_A^{(n)}$ the successive renormalized 
renormalized coupling, one obtains the product structure 
\begin{eqnarray}
V_A^{(N-1)}  =  R_{N-1} ... R_2 R_1 
\label{productA}
\end{eqnarray}
where 
\begin{eqnarray}
R_{p} &&= \frac{1}{\sqrt 2}
 \left[ \frac{\epsilon_2^{(p)}-E + V_1 (-1)^{p}}
{ \frac{\epsilon_1^{(p)}+\epsilon_2^{(p)}}{2}-E+V_1 (-1)^{p}} \right] 
\label{Rp}
\end{eqnarray}

Similarly, denoting by $\tilde \epsilon_{2}^{(n)}$ and $\tilde \epsilon_{1}^{(n)}$
the two last renormalized on-site energies after $(n-1)$ RG steps
(so that the initial on-site energies correspond to $\tilde \epsilon_{2}^{(n=1)}=\epsilon_{2^N-1}$
and $\tilde \epsilon_{1}^{(n=1)}=\epsilon_{2^N}$),
the renormalized coupling to site $B$ reads
\begin{eqnarray}
V_B^{(N-1)}  =  {\tilde R}_{N-1} ... {\tilde R}_2 {\tilde R}_1  
\label{productB}
\end{eqnarray}
where
\begin{eqnarray}
{\tilde R}_{p} &&= \frac{1}{\sqrt 2}
 \left[ \frac{{\tilde \epsilon}_2^{p}-E+V_1 (-1)^p}
{ \frac{{\tilde \epsilon}_1^{p}+{\tilde \epsilon}_2^{p}}{2}-E+V_1 (-1)^p} \right] 
\label{tildeRp}
\end{eqnarray}

After these $(N-1)$ RG steps, the renormalized system contains only two renormalized sites
$\vert 1_{(N-1)}>$ and $\vert 2_{(N-1)}>$, and the corresponding renormalized Hamiltonian reads
\begin{eqnarray}
h_{last} && = \epsilon_1^{(N-1)} \vert 1_{(N-1)}> < 1_{(N-1)} \vert 
+  \epsilon_2^{(N-1)} \vert 2_{(N-1)}> < 2_{(N-1)} \vert \nonumber \\
&& + (-1)^{N-1}V_1 \left( \vert 2_{(N-1)}> < 1_{(N-1)} \vert + \vert 1_{(N-1)}> < 2_{(N-1)} \vert \right) \nonumber \\
&& +
V_A^{(N-1)}  \left( \vert A> < 1_{(N-1)} \vert + \vert 1_{(N-1)}> < A \vert \right)
+ V_B^{(N-1)}  \left( \vert B> < 2_{(N-1)} \vert + \vert 2_{(N-1)}> < B \vert \right)
\label{HRlast}
\end{eqnarray}
The successive elimination of $\vert 1_{(N-1)}>$ and $\vert 2_{(N-1)}>$ via Aoki RG rules 
finally yields
the following renormalized coupling between the two wires ends $A$ and $B$ 
\begin{eqnarray}
V_{AB}^{eff} 
= (-1)^{N-1}V_1 \frac{ V_A^{(N-1)} V_B^{(N-1)} }
{ (E-\epsilon_1^{(N-1)})  (E-\epsilon_2^{(N-1)})  -V_1^2 }
\label{rgruleslast}
\end{eqnarray}

Since the renormalized on-site energies $\epsilon_1^{(N-1)} $ and $\epsilon_2^{(N-1)} $
have a fixed probability distribution independent of $N$, the statistical properties of the Landauer
transmission for large $N$ can be obtained from the the renormalized hoppings alone
\begin{eqnarray}
T_{N} \sim \vert V_{AB}^{eff} \vert^2  \sim  V_1^2 \vert V_A^{(N-1)} \vert^2 
\ \ \vert V_B^{(N-1)} \vert^2
\label{transmission}
\end{eqnarray}
In conclusion, the only important $N$-dependent factors are the two independent renormalized couplings 
$ V_A^{(N-1)}$ and $ V_B^{(N-1)}$ given by the products of Eqs \ref{productA} and \ref{productB}.

 \subsection{ Multifractal statistics of the transmission }

At Anderson localization transitions, the two-point transmission displays multifractal properties 
in direct correspondence with the multifractality of critical eigenstates 
\cite{janssen99,evers08,us_twopoints} : the critical probability distribution of $T_L$ over the disordered samples
takes the form
\begin{equation}
{\rm Prob}\left( T_L \sim L^{-\kappa}  \right) dT
\oppropto_{L \to \infty} L^{\Phi(\kappa) } d\kappa
\label{phikappa}
\end{equation}
and its moments involve non-trivial exponents $X(q)$
\begin{equation}
\overline{T_L^q} \sim \int d\kappa L^{\Phi(\kappa) -q \kappa }
\oppropto_{L \to \infty} L^{-X(q)}
\label{defXq}
\end{equation}
As stressed in \cite{janssen99}, the physical bound $T_L \leq 1$
on the transmission implies that the multifractal spectrum
exists only for $\kappa \geq 0$, and this termination at $\kappa=0$
 leads to a complete freezing of the moments exponents
 \begin{eqnarray}
X(q)  =X(q_{sat}) \ \ \ \ {\rm for } \ \ q \geq q_{sat}
\label{freezing}
\end{eqnarray}
at the value $q_{sat}$ where the saddle-point of the integral
of Eq. \ref{defXq} vanishes $\kappa(q \geq q_{sat})=0$.

In our present case, the moments of the transmission of Eq. \ref{transmission}
 read
\begin{eqnarray}
\overline{ T_{N}^q } \sim (V_1^2)^q \left[ \overline { \vert V_A^{(N-1)}\vert^{2q} } \right]^2
\oppropto \frac{1}{L_N^{X(q)}} = 2^{- N X(q)}
\label{tnq}
\end{eqnarray}
Using Eq. \ref{productA}, one finally obtains 
 that the exponents $X(q)$ read in terms of the variables $R_p$ of Eq. \ref{Rp}
\begin{eqnarray}
  X(q) =    
 - \oplim_{N \to \infty}  \frac{ 2 }{N  \ln 2} \ln \left[   \overline { \vert   R_{N-1} ... R_2 R_1  \vert^{2q} }  \right]
\label{xqrp}
\end{eqnarray}
The exponents $X(q)$ thus represent the 'generalized Lyapunov exponents' for the moments of products
of the variables $R_p$, that are correlated via the renormalization procedure.
Whereas for arbitrary $q$, these correlations make difficult the analytical computation of $X(q)$,
the problem becomes simple if one considers the typical behavior near $q=0$ where one has to evaluate
sums instead of products, as we now describe.

The typical transmission
\begin{eqnarray}
   T^{typ}_L \equiv e^{ \overline{ \ln T_L } } 
\label{defTtyp}
\end{eqnarray}
is expected to decay at criticality with some power-law
\begin{eqnarray}
   T^{typ}_L  
\oppropto_{L \to \infty} \frac{1}{L^{\kappa_{typ}}}
\label{defTtypcriti}
\end{eqnarray}
where $\kappa_{typ}$ is the value where $\Phi(\kappa)$ is maximum and vanishes $\Phi(\kappa_{typ})=0$.
In our present case, we may rewrite Eq. \ref{productA} as
\begin{eqnarray}
\ln \vert V_A^{(m)} \vert^2 = \sum_{n=1}^{m} \ln R_n^2
\label{vacarretypsum}
\end{eqnarray}
Using the stationary measures $C_{a,b}$ with $a_{even}=\epsilon_0+V_1$ and $a_{odd}=\epsilon_0$
for renormalized energies at even and odd steps (Eq. \ref{anoddeven}),
one obtains that $(\epsilon_1+\epsilon_2)/2$ is distributed with the same law as $\epsilon_1$ and $\epsilon_2$.
As a consequence, the average over the stationary measure yields
\begin{eqnarray}
<< \ln  R_n^2 >> = - \ln 2 
\label{rntyp}
\end{eqnarray}
and we obtain the typical decay
\begin{eqnarray}
\overline{ \ln \vert V_A^{(m)} \vert^2 }  \sim  - m \ln 2 = - \ln L_m
\label{vacarretyp}
\end{eqnarray}
in terms of the length $L_m=2^m$.
Finally, the full transmission of Eq. \ref{transmission}
has for typical behavior in $L$ 
\begin{eqnarray}
\overline{ \ln T_N }  \sim   \overline{\ln \vert V_A^{(N-1)} \vert^2}
+ \overline{\ln \vert V_B^{(N-1)} \vert^2}
= - 2 (N-1) \ln 2 
\sim  - \kappa_{typ}\ln L_N \ \ {\rm with } 
  \ \ \kappa_{typ}=2
\label{transtyp}
\end{eqnarray}
Note that this corresponds to the naive estimate $T_N \sim V_N^2 \sim 1/L_N^2$
in terms of the direct coupling $V_N$ between the sites $1$ and $2^N$.
In the next section, we explain the consequences for the multifractal statistics of eigenstates.

 \section{ Multifractal statistics of eigenstates at criticality $\gamma_c=-1/2$ } 

\label{sec_falpha}

One of the most important property of Anderson localization transitions
is that critical eigenfunctions are described by
 a singularity spectrum $f(\alpha)$ defined as follows
(for more details see for instance the review \cite{mirlinrevue}):
in a sample of size $L^d$,  the number ${\cal N}_L(\alpha)$
of points $\vec r$ where the weight $\vert \psi(\vec r)\vert^2$
scales as $L^{-\alpha}$ behaves as 
\begin{eqnarray}
{\cal N}_L(\alpha) \oppropto_{L \to \infty} L^{f(\alpha)}
\label{nlalpha}
\end{eqnarray}
The inverse participation ratios (I.P.R.s) can be then rewritten
as an integral over $\alpha$
\begin{equation}
Y_q(L)  \equiv \int_{L^d} d^d { \vec r}  \vert \psi (\vec r) \vert^{2q}
\simeq \int d\alpha \ L^{f(\alpha)} 
\ L^{- q \alpha} \opsimeq_{L \to \infty} 
L^{ - \tau(q) }
\label{ipr}
\end{equation}
where $\tau(q) $ can be obtained via the Legendre transform formula
\begin{eqnarray}
- \tau(q) = {\rm max}_{\alpha} \left[ f(\alpha) - q \alpha \right]
\label{legendre}
\end{eqnarray}

 \subsection{ Strong disorder regime } 

As recalled in the introduction, the 'strong multifractality' regime
has been first studied via the powerful Levitov renormalization method \cite{levitov}
(see also the reformulation as
some type of 'virial expansion' in Refs \cite{oleg1,oleg2,oleg3,oleg4}) :
the obtained leading order result is the same for
the PRBM model \cite{mirlin_evers} and for the 
'ultrametric random matrices ensemble' \cite{fyodorov,EBetOG},
and can also be derived via a simpler direct first-order perturbation theory for eigenstates 
\cite{us_strong}. For our present model, we may directly apply the analysis of \cite{us_strong},
where the only important factor is the two-point hopping (see section 2.1.2. in \cite{us_strong}).
We thus conclude that our present model is described by the same 'universal' strong multifractality
regime for large $W$
\begin{eqnarray}
 f^{SM}(\alpha) = \frac{ \alpha }{2} \ \  {\rm for } \ \ 0 \leq \alpha \leq 2
\label{strongmultif}
\end{eqnarray}
or equivalently for the exponents $\tau(q)$
\begin{eqnarray}
\tau^{SM}(q) && = d (2q-1) \ \ {\rm for } \ \ q<\frac{1}{2} \nonumber \\
\tau^{SM}(q) && = 0 \ \ {\rm for } \ \ q>\frac{1}{2}
\label{sm}
\end{eqnarray}
The first correction in $1/W$
discussed in \cite{mirlin_evers,fyodorov,us_strong,EBetOG}
involves an explicit perturbative universal correction for $\tau(q)$
in the region $q>1/2$, that should hold also in our present model.
 But then to obtain the perturbative correction 
for $\tau(q)$ for $q<1/2$, and in particular
the behavior of $f(\alpha)$ near the typical value ($\alpha_{typ}=\alpha(q \to 0)$), 
one has to rely on the 'multifractal symmetry'
(see more details in section \ref{numemultifsym}). 
Since it is not clear to us whether this symmetry is satisfied in the present
Dyson model, 
we cannot conclude that the whole first order correction for all values $(\alpha,q)$
is the same as in the PRBM model \cite{mirlin_evers} and in the 
'ultrametric random matrices ensemble' \cite{fyodorov,EBetOG}.
In particular, in these two models, the typical value $\alpha_{typ}$ is known to
move from $\alpha_{typ}^{SM}=2$ at first order (i.e. $2-\alpha_{typ} \propto 1/W$), 
whereas in the present model,
it seems to us that the typical value $\alpha_{typ}$ remains frozen to the value $\alpha_{typ}^{SM}=2$ independently of the disorder $W$, as we now discuss.

 \subsection{ Relation with the multifractal statistics of the transmission }

The multifractal spectrum $\Phi(\kappa)$ of Eq. \ref{phikappa}
is expected to be directly related
to the singularity spectrum $f(\alpha)$ of Eq. \ref{nlalpha}
 via \cite{janssen99,evers08,us_twopoints}
\begin{eqnarray}
\Phi(\kappa \geq 0) = 2 \left[ f( \alpha= d+ \frac{\kappa}{2}   ) -d \right]
\label{resphikappa}
\end{eqnarray}
In particular, the typical exponents $\kappa_{typ}$ and $\alpha_{typ} $
where $\Phi(\kappa_{typ} )=0 $ and $f(\alpha_{typ})=d $ are related by  \cite{janssen99,evers08,us_twopoints}
\begin{eqnarray}
\alpha_{typ}= d+ \frac{\kappa_{typ}}{2} 
\label{resphikappatyp}
\end{eqnarray}
In our present case where $d=1$, the result $\kappa_{typ}=2 $ of Eq. \ref{transtyp}
yields
\begin{eqnarray}
\alpha_{typ}=2 
\label{alphatyp}
\end{eqnarray}
independently of the disorder strength $W$.
This behavior is very anomalous with respect to other models where 
the typical exponent usually moves continuously between the weak-disorder value
$\alpha_{typ}(W \to 0)=d$ and the strong disorder value $\alpha_{typ}(W \to \infty)=2 d$.

In terms of the moments exponents, Eq. \ref{resphikappa} is equivalent in the region $q \leq q_{sat}$
to  \cite{janssen99,evers08,us_twopoints}
 \begin{eqnarray}
X(q)  = 2 \left[ \tau(q)-d (q-1) \right]
\label{Xqtauq}
\end{eqnarray}

 \subsection{ Numerical results for the multifractal spectrum as a function of $W$  } 

\label{numemultif}

\begin{figure}[htbp]
 \includegraphics[height=6cm]{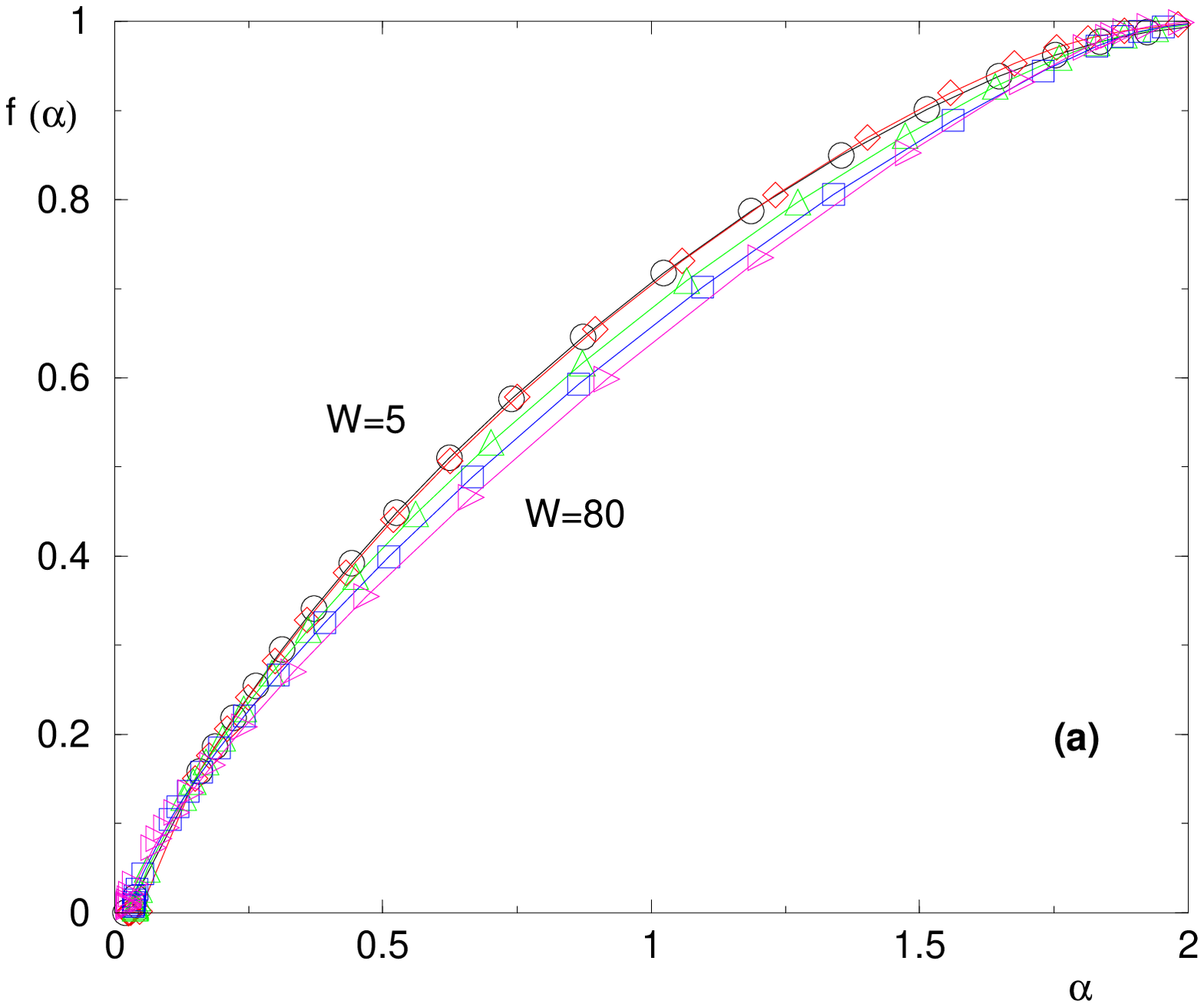}
\hspace{2cm}
 \includegraphics[height=6cm]{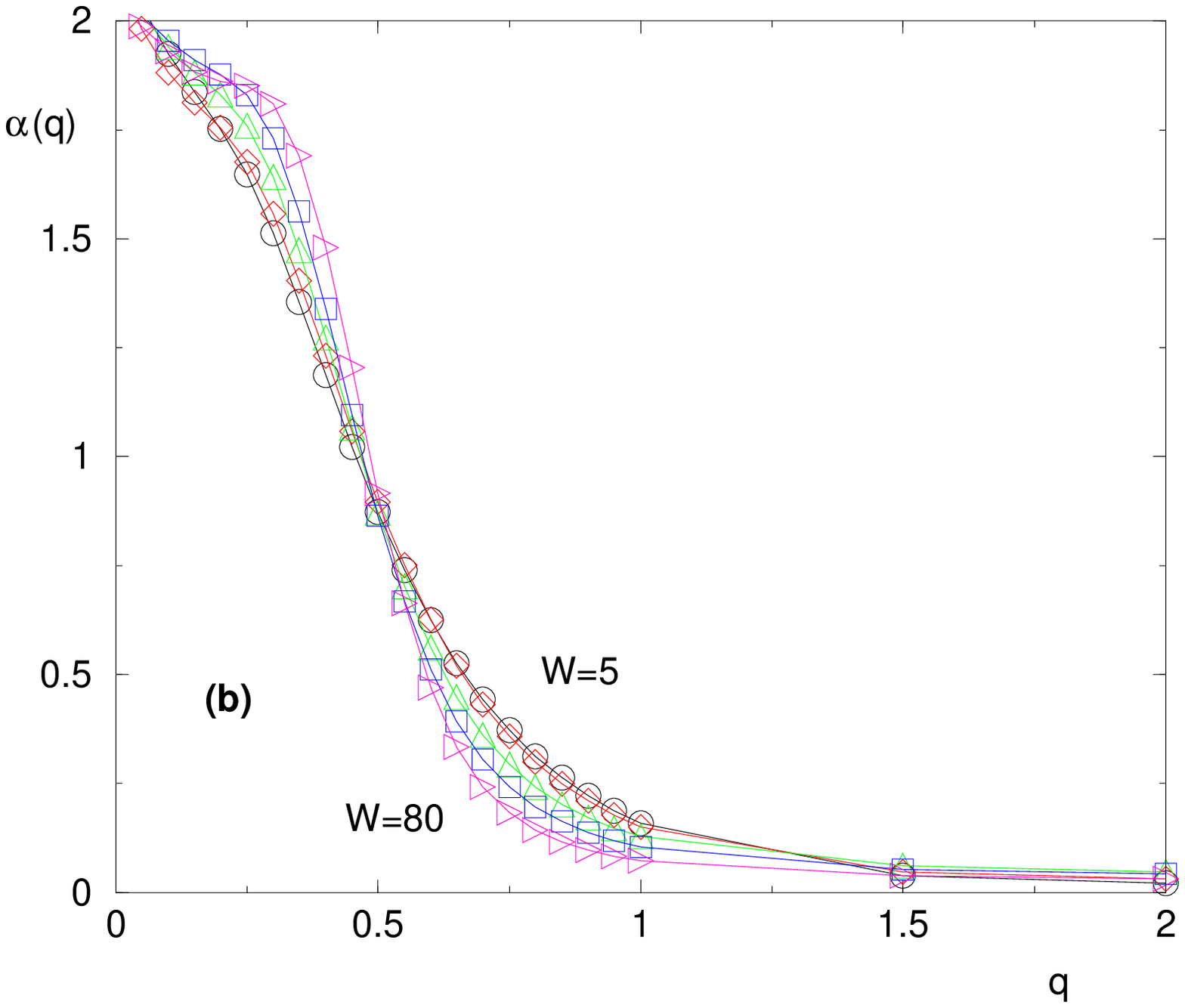}
\vspace{1cm}
\caption{  Box disorder :  multifractal statistics of eigenfunctions
for the disorder strengths  $W=5,10,20,40,80$
(a) singularity spectrum $f(\alpha)$
(b) corresponding $\alpha(q)$ }
\label{figfalphabox}
\end{figure}

\begin{figure}[htbp]
 \includegraphics[height=6cm]{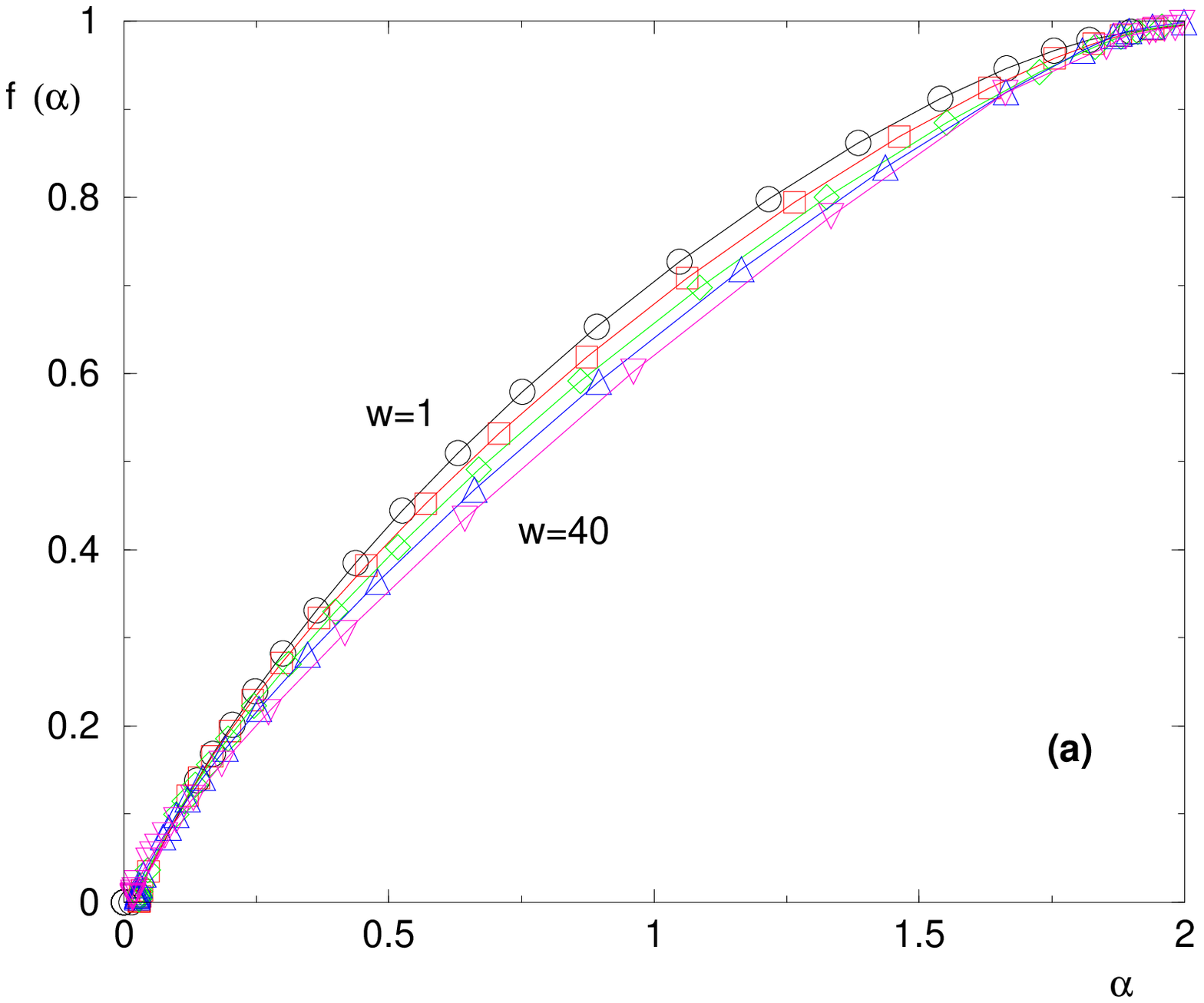}
\hspace{2cm}
 \includegraphics[height=6cm]{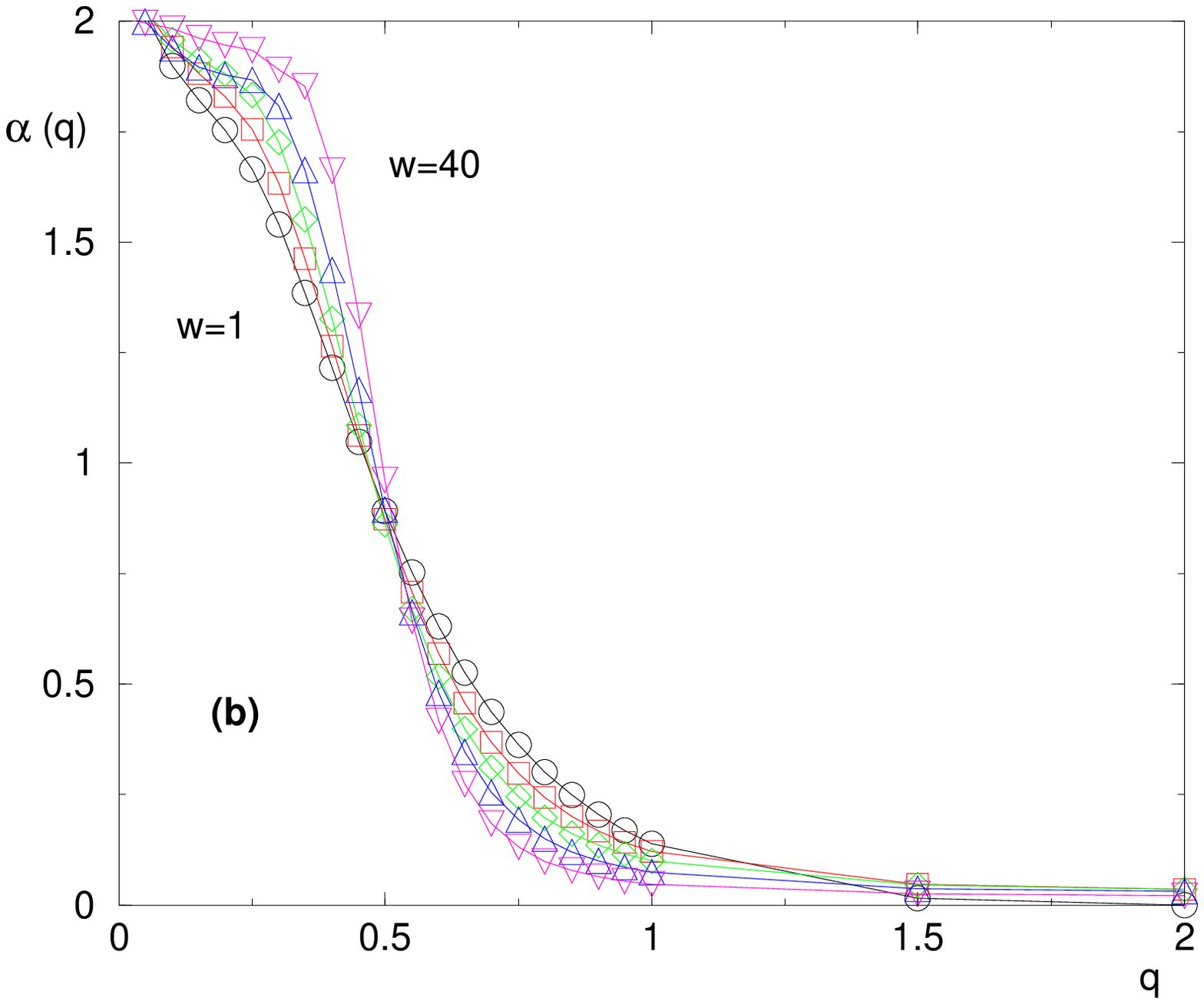}
\vspace{1cm}
\caption{  Cauchy disorder : multifractal statistics of eigenfunctions for the disorder strengths  $W=1,5,10,20,40$
(a) singularity spectrum $f(\alpha)$
(b) corresponding $\alpha(q)$ }
\label{figfalphacauchy}
\end{figure}

We have studied via exact diagonalization 
disordered samples of sizes $L_N=2^N$ (with $N=7,8,9,10,11,12$ generations) 
 with the following corresponding numbers $n_s(L_N)$
of disordered samples
\begin{eqnarray}
N && = 128 ; 256 ; 512 ; 1024 ; 2048 ; 4096 \nonumber \\
n_s(N) && = 51.10^5 ; 11.10^5 ; 215.10^3 ; 326.10^2 ; 3.10^3 ; 780
\label{numerics} 
\end{eqnarray}
In each disordered sample, we have analyzed the fraction $1/8$ of the eigenstates
at the center of the spectrum (we have checked that the density of states is nearly constant
in this region and that the corresponding eigenstates have the same statistics of I.P.R.).
The multifractal spectrum $f(\alpha)$ is then obtained parametrically in $q$
via the standard method of Ref \cite{Chh} (see more details in Appendix B of \cite{us_cayleymultif}).
We have chosen the values $\epsilon_0=0$ and $V_1=1$.

\subsubsection{ Results for the box distribution  }

We show on Fig. \ref{figfalphabox} our numerical results concerning the singularity spectrum $f(\alpha)$
for the disorder strengths $W=5,10,20,40,80$. 
At strong disorder $W=80$, the multifractal spectrum is very close to the strong multifractal
universal result of Eq. \ref{strongmultif}  as expected.
As the disorder strength $W$ becomes smaller, the multifractal spectrum becomes more curved
but keeps a 'strong multifractal' character, with a typical value 
remaining at the maximal value $ \alpha_{typ} \simeq 2 $, and an
almost vanishing minimal value $\alpha_{min} \simeq 0$. The inhomogeneity of eigenfunctions is thus always very strong.

For the box disorder below $W=5$, the density of states
tend to break into two bands around the two pure delta peaks of Eq. \ref{dospurecriti} that are separated here by $\epsilon_{even}^- -\epsilon_{odd}^- = 3 V_1=3$. As a consequence, the choice to work around $E=0$ is not appropriate anymore at weak disorder, and we discuss this limit in section \ref{weak}.

\subsubsection{ Results for the Cauchy distribution  }

We show on Fig. \ref{figfalphacauchy} our numerical results concerning the singularity spectrum $f(\alpha)$ for the disorder strengths $W=1,5,10,20,40$ : the results are qualitatively similar to
the box disorder case.

Note that for the Cauchy disorder, the density of states can be computed exactly from the 
pure density of states \cite{llyod}
\begin{eqnarray}
\rho(E) = \int dE' \rho^{pure}(E') \frac{W}{\pi \left[ (E-E')^2+W^2 \right] }
\label{dosllyod} 
\end{eqnarray}
i.e. in our present case with the form of Eq. \ref{dospurecriti} for $\rho^{pure}(E') $,
we obtain
\begin{eqnarray}
\rho(E) =  \frac{2}{3} \frac{W}{\pi \left[ (E- \epsilon_-^{odd})^2+W^2 \right] } + \frac{1}{3} 
\frac{W}{\pi \left[ (E- \epsilon_-^{even})^2+W^2 \right] }
\label{doscriti} 
\end{eqnarray}
with the numerical values  $\epsilon_-^{odd}=-1$ and $\epsilon_-^{even}=2$ (Eqs \ref{emdegecriti} 
with $\epsilon_0=0$ and $V_1=1$)

\subsubsection{ Test of the symmetry $f(2d-\alpha)=f(\alpha)+d-\alpha $ }

\label{numemultifsym}

\begin{figure}[htbp]
 \includegraphics[height=6cm]{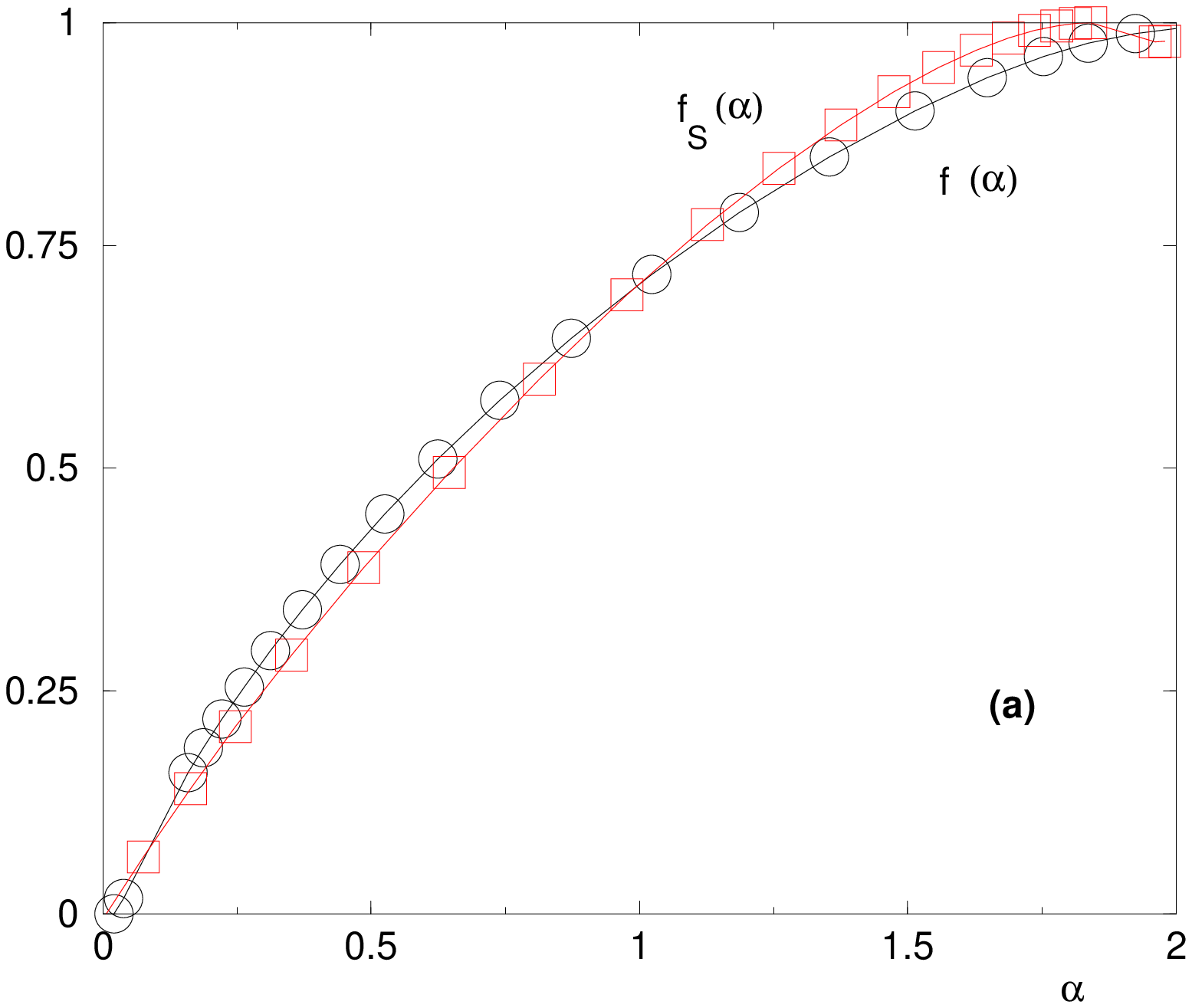}
\hspace{2cm}
 \includegraphics[height=6cm]{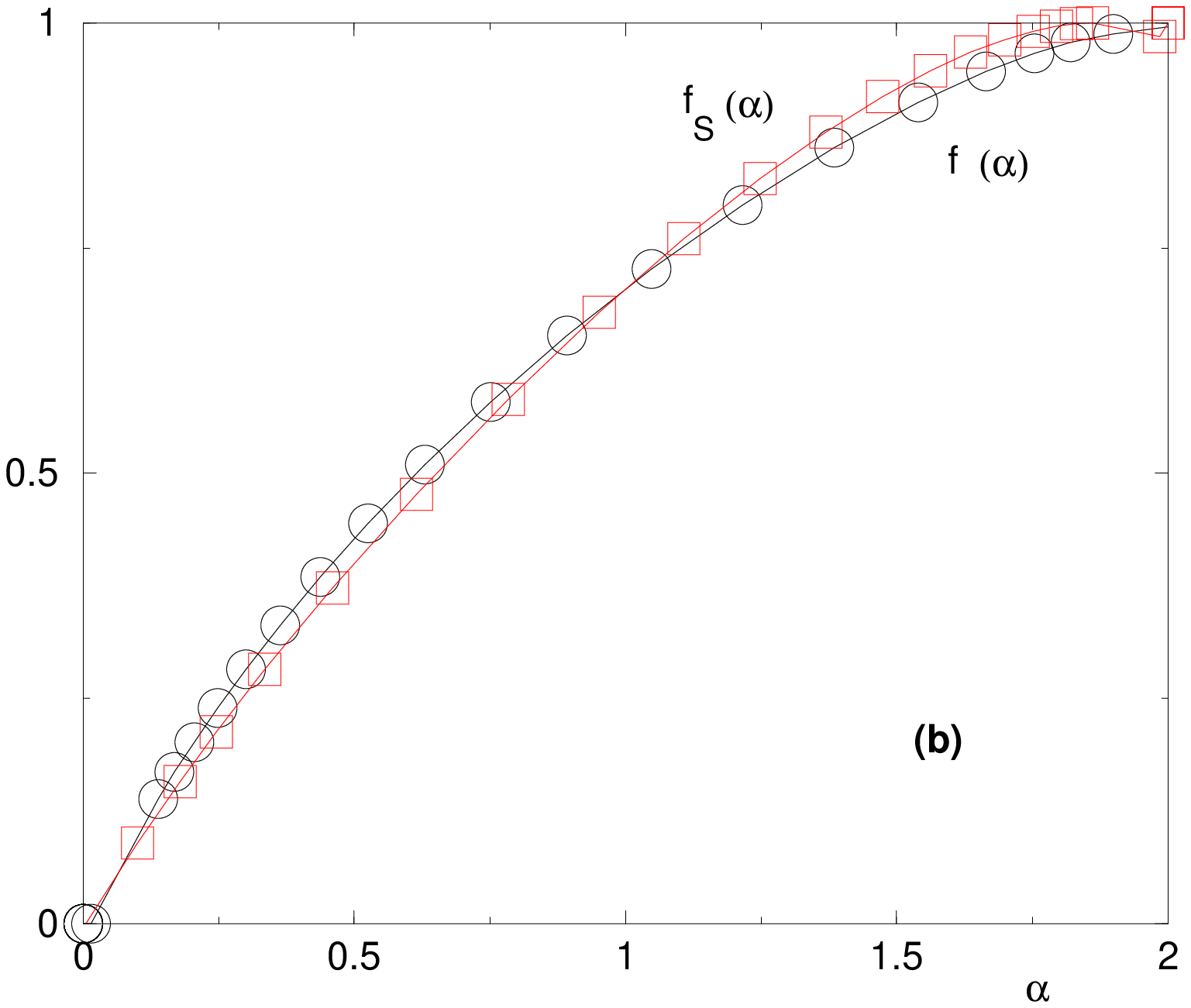}
\vspace{1cm}
\caption{  Test of the symmetry of Eq. \ref{symfaanderson} : we compare
the singularity spectrum $f(\alpha)$ and the function $f_S(\alpha)\equiv f(2-\alpha)-(1-\alpha)$ 
(a) for the box disorder with $W=5$
(b) for the Cauchy disorder with $W=1$ }
\label{figtestsym}
\end{figure}

For any Anderson transition in the so-called 
'conventional symmetry classes' \cite{mirlinrevue},
 Mirlin, Fyodorov, Mildenberger and Evers \cite{mirlin06}
have proposed that the singularity spectrum $f(\alpha)$ of critical
eigenfunctions satisfies the remarkable exact symmetry 
\begin{eqnarray}
f(2d-\alpha)=f(\alpha)+d-\alpha
\label{symfaanderson}
\end{eqnarray}
that relates the regions $\alpha \leq d$ and $\alpha \geq d$.
Further discussions can be found in \cite{mirlinrevue,vasquez,us_sym}.
In terms of the $q$-variable, the symmetry 
with respect to the value $\alpha_s=d$ becomes 
a symmetry with respect to the value $q_s=1/2$ \cite{mirlin06},
so that one should observe the following fixed point
\begin{eqnarray}
\alpha \left(q=\frac{1}{2}\right)=1
\label{alphaqdemi}
\end{eqnarray}
On Fig. \ref{figfalphabox} (b) for the box disorder case, and on Fig. \ref{figfalphacauchy}
for the Cauchy disorder case, we find that the curves $\alpha_W(q)$ for various disorder strength $W$
cross near the point of Eq. \ref{alphaqdemi}.
To test more directly the symmetry of Eq. \ref{symfaanderson} with $d=1$, 
we have plotted $f(\alpha)$ and $f_S(\alpha)\equiv f(2-\alpha)-(1-\alpha)$
together on Fig. \ref{figtestsym}, for the box disorder at $W=5$ and for the Cauchy disorder at $W=1$ :
the difference between the two remains within our numerical errors.
For larger $W$, the difference become smaller, as could be expected since the
strong multifractality limit of Eq. \ref{strongmultif}  satisfies the symmetry exactly.
In summary, since the deviations with respect to 'strong multifractal limit'
are small, the deviations from the symmetry are also small,
and it seems difficult to obtain a clear numerical conclusion.
On the other hand, the discussion after Eq. \ref{sm} suggests that the multifractal
symmetry is not compatible with our statement of Eq. \ref{alphatyp}
concerning the fixed value of the typical exponent $\alpha_{typ}=2$ 
independently of the disorder strength $W$.
The clarification of this point goes beyond the present work.

 \section{ Compressibility of energy levels at criticality $\gamma_c=-1/2$ } 

\label{sec_eigen}

\begin{figure}[htbp]
 \includegraphics[height=6cm]{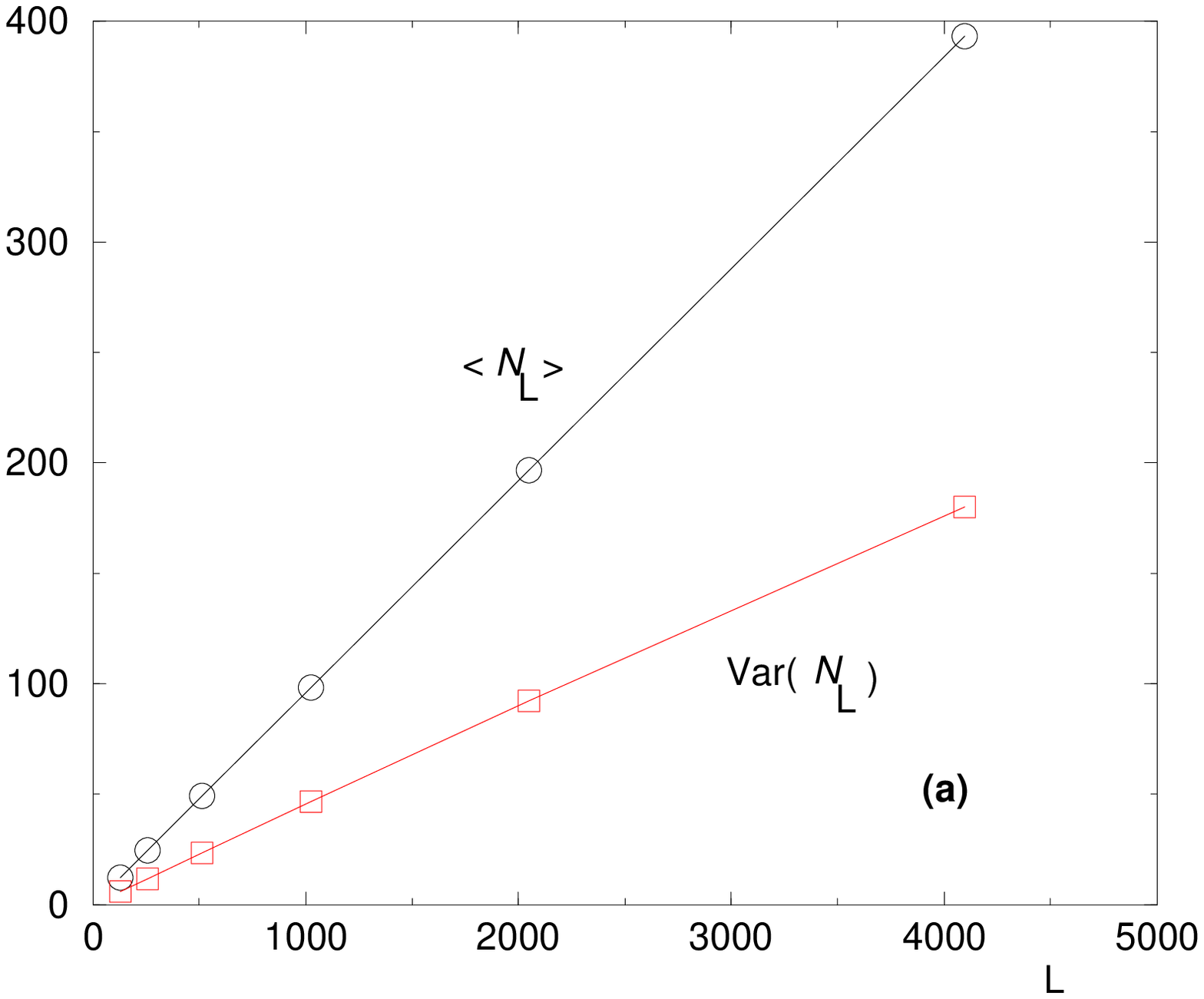}
\hspace{2cm}
 \includegraphics[height=6cm]{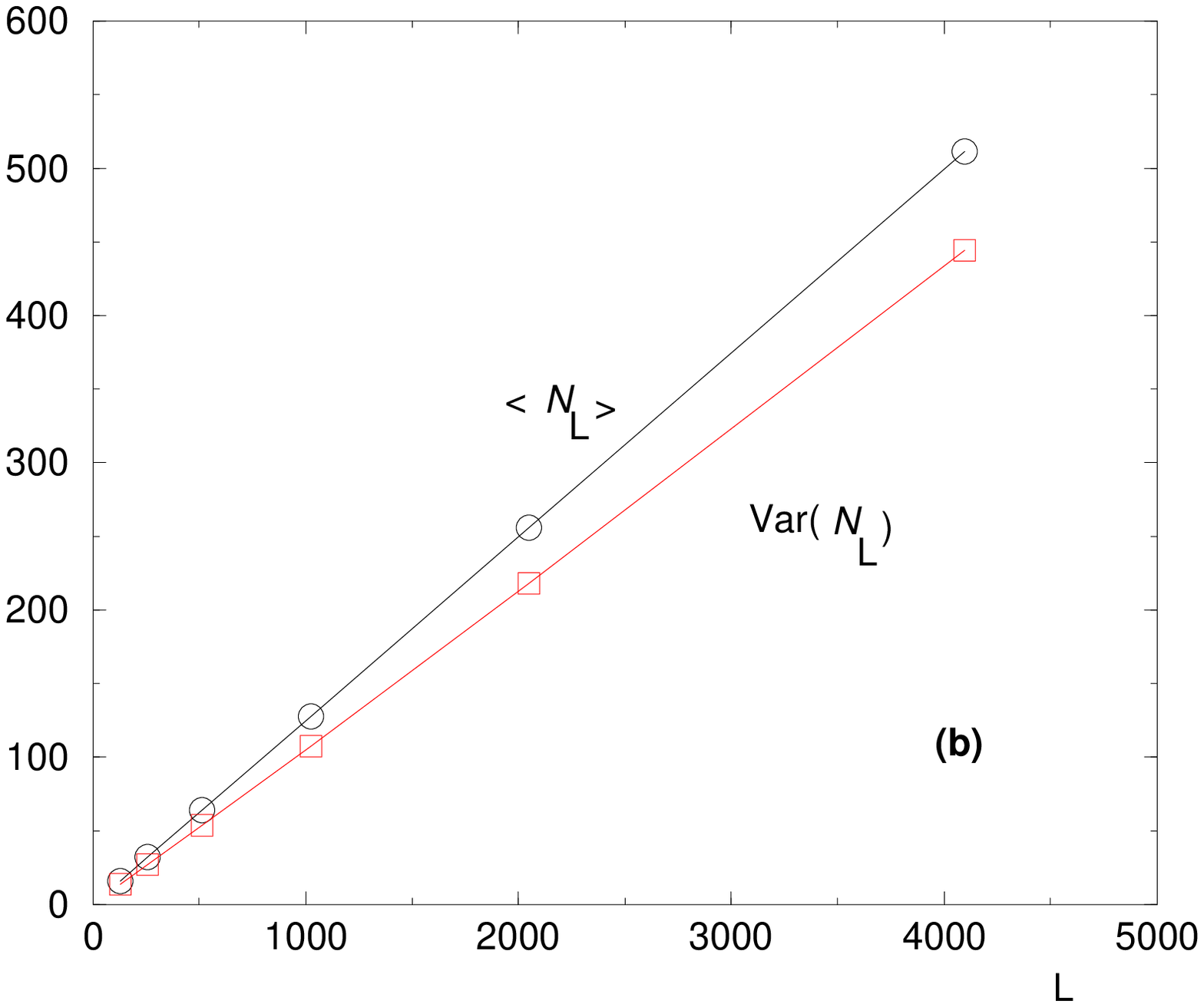}
\vspace{1cm}
\caption{ Box disorder :  statistics of the number ${\cal N}_L$ of eigenvalues 
  within a fixed interval $I_{\Delta E} =[-\Delta E/2,+\Delta E/2]$ as a function of the system size $L$.
Both the averaged number and the variance grow linearly in $L$, the ratio of the two slopes yields
the compressibility $\chi$
(a) Case $W=5$ : we measure $\chi \simeq 0.45$
(b) Case $W=80$ : we measure $\chi \simeq 0.87$ }
\label{figchibox}
\end{figure}

Another important property of Anderson localization transitions
is that the statistics of eigenvalues is neither 'Poisson' (as in the localized phase)
nor 'Random Matrix' (as in the delocalized phase) but 'intermediate'
(see for instance \cite{eigenps93,eigenps97,cuevas} and references therein ).
A convenient parameter is the level compressibility $\chi$,
which is found to satisfy $0<\chi<1$ at Anderson transitions
(see for instance \cite{EBetOG} and references therein),
whereas delocalized states are characterized by
 $\chi_{deloc}=0$ and localized states by $\chi_{loc}=1$.

We have studied via exact diagonalization 
disordered samples of sizes $L_N=2^N$ (with $N=7,8,9,10,11,12$ generations) 
 with the following corresponding numbers $n_s(L_N)$
of disordered samples
\begin{eqnarray}
L_N && = 128 ; 256 ; 512 ; 1024 ; 2048 ; 4096 \nonumber \\
n_s(L_N) && = 41.10^6 ; 915.10^4 ; 17.10^5 ; 19.10^4 ; 12.10^3 ; 48.10^2
\label{numericseigen} 
\end{eqnarray}
In each disordered sample, we have analyzed the number of eigenvalues
within a fixed interval $I_{\Delta E} =[-\Delta E/2,+\Delta E/2]$
where the density of states is nearly constant.
The averaged number of eigenstates in this interval 
for disordered samples of size $L$ scales as
\begin{eqnarray}
< {\cal N}_L  > \opsimeq_{L \to +\infty}  L \rho(0) \Delta E 
\label{numberav} 
\end{eqnarray}
The compressibility is then defined by the ratio between the variance and the averaged number
\begin{eqnarray}
\frac{Var ({\cal N}_L) }{< {\cal N}_L  >}  \opsimeq_{L \to +\infty}  \chi 
\label{chi} 
\end{eqnarray}

\subsection{ Results for the box distribution  }

On Fig. \ref{figchibox}, we show for the two cases $W=5$ and $W=80$ the linear behavior in $L$
of the averaged number of eigenstates $< {\cal N}_L  > $ and of the variance $Var ({\cal N}_L) $.
Our final results concerning the compressibility as a function of the disorder strength $W$ are
\begin{eqnarray}
 W  && =     5 ,  10 ,  20 , 40 , 80  \nonumber \\
 \chi(W)  && =  0.45 ,  0.45 , 0.62 , 0.79 , 0.87 
\label{chibox} 
\end{eqnarray}
We find that our results for $W=5$ and $W=10$ are nearly the same, as already found for the singularity spectrum $f(\alpha)$. Then the compressibility $\chi$ grows with $W$ as expected, and goes to $1$ in the strong disorder limit $W \to +\infty$.
A direct relation $\chi+D_1/d=1$
between the compressibility $\chi$ of energy levels, and the information dimension
$D_1=\alpha_{q=1}$ of eigenfunctions has been recently conjectured and checked in various
models \cite{EBetOG} : for the present model, our numerical results do not seem compatible
with this relation. For instance at $W=5$,  the measured
information dimension $D_1=\alpha_1 \simeq 0.15$ would correspond 
via the conjectured relation in $d=1$ to $\chi'=1-D_1 \simeq 0.85$,
whereas we measure the compressibility $\chi \simeq 0.45$.
Our conclusion is thus that the relation $\chi+D_1/d=1$ conjectured in \cite{EBetOG}
probably needs some hypothesis that is not satisfied by the present model.

\subsection{ Results for the Cauchy distribution }

For the Cauchy disorder, our final results for the compressibility as a function of the disorder strength $W$ read
\begin{eqnarray}
 W  && =     1, 5 ,  10  , 40   \nonumber \\
 \chi(W)  && =  0.60 ,  0.64 , 0.77 , 0.96 
\label{chicauchy} 
\end{eqnarray}
Here for $W=40$, the obtained compressibility is very near the Poisson value $1$.

\section{ Anomalous weak-disorder limit }

\label{weak}

\subsection{ Numerical results within the lower band }

In the previous sections \ref{numemultif} and \ref{sec_eigen},
we have shown numerical results for the multifractal spectrum and the compressibility at the center of the band $E=0$ for finite disorder $W$.
However in the weak-disorder region, the density of states
tend to break into two bands around the two pure delta peaks of Eq. \ref{dospurecriti} that are separated here by $\epsilon_{even}^- -\epsilon_{odd}^- = 3 V_1=3$. As a consequence, the choice to work around $E=0$ is not appropriate anymore at weak disorder, but one can instead work in one of the two sub-bands.
We have chosen to study the statistical properties of the eigenvalues and eigenvectors of a fraction $1/8$ of the states of the lower sub-band
(after checking that the density of states was nearly constant in this region).

For the Box distribution, we 
find that the compressibility $\chi$ takes the values
\begin{eqnarray}
 W  && =     0.1  ,  1  \nonumber \\
 \chi^{Box}(W)  && \simeq  0.41 ,  0.42  
\label{chiboxweak} 
\end{eqnarray}
For the Cauchy distribution, we actually find the same limiting value
\begin{eqnarray}
 \chi^{Cauchy}(W=0.01)  \simeq  0.41   
\label{chicauchyweak} 
\end{eqnarray}
These numerical results indicate
 that the compressibility remains finite in the weak disorder 
regime $W \to 0^+$,
whereas in other models, it vanishes smoothly in the disorder strength
to recover the 'Random Matrix' value $\chi=0$.

We have also analyzed the multifractal spectrum of eigenstates
for the values of $W$ given above : they keep a 'strong multifractality'
character, with a typical value around $\alpha_{typ} \simeq 2$
and a minimal value $\alpha_{min}$ near zero.

\subsection{ Discussion }

The numerical results obtained in a given sub-band of the pure model
indicate that the 'weak-disorder' regime is very anomalous
in the Dyson hierarchical model. 
Indeed  in usual models characterized by a non-degenerate continuum of plane waves, the non-degenerate perturbation theory yields the universal first order Gaussian correction in $q(q-1)$ to the multifractal spectrum (see \cite{mirlinrevue,fyodorovbis} for more details on this 'weak multifractality' regime),
and the compressibility is perturbatively close to the 'Random Matrix' value $\chi=0$. Here these generic results do not apply,
because the pure model is extremely degenerate (Eq. \ref{dospurecriti}),
so that one should diagonalize the perturbation in each extensively degenerate
subspace of the pure model. In some sense, this means that the perturbation
is never 'weak', since there is no energy scale associated to the
pure model in a given delta-peak.

\section{ Conclusion } 

\label{sec_conclusion}

In this paper, we have described how the 
 Dyson hierarchical model for Anderson localization, containing non-random hierarchical hoppings
and random on-site energies, can reach an Anderson localization critical point presenting multifractal eigenfunctions and intermediate spectral statistics, provided one 
introduces alternating signs in the hoppings along the hierarchy 
(instead of choosing all hoppings of the same sign as had been done up to now 
in the mathematical literature \cite{bovier,molchanov,krit,kuttruf}).
This model is somewhat simpler than the 'ultrametric random matrices ensemble' considered by physicists \cite{fyodorov,EBetOG,fyodorovbis}, because here 
it is directly the matrix of non-random hoppings in each sample
that presents a hierarchical block structure,
whereas in References \cite{fyodorov,EBetOG,fyodorovbis}, it is only 
the matrix of the variances of the random hoppings that presents a hierarchical block structure.
In particular, we have obtained exact renormalization equations for some 
observables, like the renormalized on-site energies or the renormalized couplings to exterior wires.
For the renormalized  on-site energies, we have showed that the Cauchy distributions 
are exact fixed points. From the renormalized couplings to exterior wires,
we have obtained that the typical exponent of eigenfunctions
is always $\alpha_{typ}=2$ independently of the disorder strength, in agreement with
our numerical exact diagonalization results for the box distribution and for the Cauchy distribution
of the random on-site energies. 
We have also explained how this model has the same universal 'strong multifractality' regime
in the limit of infinite disorder strength $W \to +\infty$ as in other models. The big difference with other
models is however that the singularity spectrum $f(\alpha)$ keeps 
for finite disorder a 'strong multifractal' character with very inhomogenous eigenfunctions,
instead of flowing towards a 'weak multifractality regime'. 
The absence of this 'weak multifractality regime' comes from the anomalous pure spectrum of this hierarchical tree structure, with two extensively degenerate delta peaks (instead of some continuum corresponding to plane waves).

We hope that the present work will stimulate further work in the
renormalization analysis of this critical model, 
or in the 'critical ultrametric ensemble'
which has only be studied via perturbation theory or numerics up to now
\cite{fyodorov,EBetOG,fyodorovbis}, since the main motivation
to introduce Dyson hierarchical models
is usually to obtain exact renormalization equations. A further goal is to better understand multifractality
via renormalization in other non-hierarchical models, as already discussed in Refs \cite{us_sym,us_travel}.

\end{document}